\documentclass[12pt]{article}
\usepackage{amsfonts,epsfig,latexsym,amscd,amsmath,theorem,mathrsfs,psfrag}
\usepackage{color}
\newcommand{\ignore}[1]{}

\newcommand{\R}{{\mathbb{R}}}
\newcommand{\Z}{{\mathbb{Z}}}

\newcommand{\C}{{\mathbb{C}}}

\newcommand{\beq}{\begin{equation}}
\newcommand{\eeq}{\end{equation}}
\newcommand{\bea}{\begin{eqnarray}}
\newcommand{\eea}{\end{eqnarray}}
\newcommand{\ben}{\begin{eqnarray*}}
\newcommand{\een}{\end{eqnarray*}}
\newcommand{\ra}{\rightarrow}

\newcommand{\cd}{\partial}
\newcommand{\wt}{\widetilde}

\def \d{\mathrm{d}}

\newcommand{\ip}[1]{\langle #1 \rangle}

\newcommand{\jsgdohk}{0.4cm}

\theoremstyle{plain}

{\theorembodyfont{\rmfamily}

}
\newcommand{\news}{\setcounter{equation}{0}}

\renewcommand{\phi}{\varphi}

\begin{document}

\title{Isospinning hopfions}
\author{Derek Harland$^{1}$,
Juha J\"aykk\"a$^{2}$,
Yakov Shnir$^{3}$ and Martin Speight$^{3}$\\
\small 
$^{1}$Department of Mathematical Sciences, Loughborough University \vspace*{-\jsgdohk}\\
\small Loughborough LE11 3TU, England \\ \vspace*{-\jsgdohk}
\small $^{2}$Nordita,
KTH Royal Institute of Technology and Stockholm University\\ \small
Roslagstullsbacken 23,
SE-106 91 Stockholm,
Sweden\\
\small $^{3}$School of Mathematics, University of Leeds \vspace*{-\jsgdohk}\\
\small Leeds LS2 9JT, England}

\ignore{
\thanks{E-mail: {\tt d.g.harland@lboro.ac.uk}}\\Department of Mathematical Sciences, Loughborough University\\ Loughborough LE11 3TU, England \\
Juha J\"aykk\"a\thanks{E-mail: {\tt }}\\juha's address \\
Yakov Shnir\thanks{E-mail: {\tt shnir@maths.tcd.ie}}\,  and Martin Speight\thanks{E-mail: {\tt speight@maths.leeds.ac.uk}}\\
School of Mathematics, University of Leeds\\
Leeds LS2 9JT, England}

\date{}
\maketitle

\begin{abstract}
The problem of constructing internally rotating solitons of fixed
angular frequency $\omega$ in the Faddeev-Skyrme
model is reformulated as a variational problem for an energy-like functional,
called pseudoenergy, which depends parametrically on $\omega$.
This problem is solved numerically using a gradient descent method, without 
imposing any spatial symmetries on the solitons, and the
dependence of the solitons' energy on $\omega$, and on
their conserved total isospin $J$, studied. 
It is found that, generically,
the shape of a soliton is independent of $\omega$, and that its size grows
monotonically 
with $\omega$. A simple elastic rod model of time-dependent hopfions is
developed which, despite having only one free parameter, accounts well
for most of the numerical results.
\end{abstract}

\noindent
PACS classification numbers: 05.45.Yv, 03.50.-z\newline

\maketitle

\section{Introduction}
\label{sec:intro}
\news

Many field theories of interest in fundamental physics support topological 
solitons -- spatially localized, stable lumps of energy whose strongly 
particle-like characteristics make them natural theoretical models of
elementary particles. Perhaps the best developed model from this viewpoint is 
the Skyrme model, whose solitons are posited to model atomic nuclei. 
It is of fundamental importance in this context 
that individual solitons possess both
rotational and internal rotational (or {\em isorotational}) degrees of freedom.
In the Skyrme model, the rotational degrees of freedom account, 
after quantization, for the spin of atomic nuclei, while the 
isorotational degrees of freedom account, roughly speaking, for their 
difference in ``flavour''. For example, both protons and neutrons are
modelled by degree one solitons, the only difference being in the sense
in which they are (internally) spinning. In practice, finding even static
classical solitons is a significant computational challenge, so it is 
not surprising that almost all studies of (iso)spinning solitons work within
a rigid body type approximation: the (iso)spinning soliton is assumed
to have precisely the same shape as the static soliton, but with its
spatial and internal orientation allowed to vary in time
\cite{adknapwit,kruspe}. It has long
been recognized  
that this is not a very satisfactory approximation, and
various attempts have been made to improve on it
\cite{sch,leemansch}. With a few exceptions \cite{acuhalnorshn,batkrusut}, 
these 
have fallen short of numerically solving the full field equations to
construct genuine (iso)spinning solutions.

In this paper we construct genuine isospinning (i.e.\ internally rotating)
soliton solutions in the Faddeev-Skyrme model \cite{fadnie}. 
Such solitons are conventionally
called hopfions, since they are classified topologically by their Hopf
degree (an integer-valued topological invariant). 
Explicitly, our field $\phi$ is $S^2$ valued, lives in $\R^{3+1}$, and has
Lagrangian density
\beq
\label{define_the_model_in_the_intro}
{\cal L}=\frac12\cd_\mu\phi\cdot\cd^\mu\phi
-\frac14(\cd_\mu\phi\times\cd_\nu\phi)\cdot(\cd^\mu\phi\times\cd^\nu\phi)
-\frac12\mu^2(1-\phi_3^2).
\eeq
This model is ideal
for our purposes because it is three-dimensional and has a rather rich
spectrum of static soliton solutions, which should be understood as
(possibly linked or self-knotted) string-like objects, but has only one
internal rotational degree of freedom, because the potential
 breaks the internal rotational symmetry group
to $SO(2)$. 
(Of course, one needs a potential term in ${\cal L}$ in order to
allow any isospinning solutions at all.)
Hence one can ask unambiguously 
 ``what is the degree $Q$ isospinning 
hopfion of
given angular frequency $\omega\in\R$?'' For a given degree $Q$, there
are usually several different stable static soliton solutions of rather
similar energy (the number of solutions seems to grow with $Q$; see
\cite{fos} for a study of static hopfions in a model
with potential). 
It is interesting to discover how the energy of
such solitons varies as their angular frequency (or their conserved isospin)
changes. We will find several examples where these energy curves cross,
so that the most energetically favourable shape of the 
hopfion for a given degree 
$Q$ changes when one isospins them fast enough.

Our method is to use the Principle of Symmetric Criticality to reduce the
problem of finding isospinning solitons of angular frequency $\omega$
to a static ``energy'' minimization problem, where the ``energy'' to be 
minimized (more properly called pseudo-energy) $F_\omega$ depends parametrically
on $\omega$. We solve this problem numerically by a standard gradient descent
method. Starting at $\omega=0$ (where the solitons are already well understood),
we gradually increment $\omega$ and reminimize, in this way constructing curves
of isospinning solitons, parametrized by $\omega$. This method allows us
to construct isospinning solitons without imposing axial symmetry, in
contrast to previous studies which have used simulated annealing
\cite{batkrusut} or direct solution of the field equations
\cite{acuhalnorshn}. Since very few static Hopf solitons are axially
symmetric, this is a significant advantage.

We expect that at least one 
 minimizer of $F_\omega$ exists for each $Q$ whilever
$\omega<
\min\{\omega_1,\omega_2\}$, where $\omega_1$ is determined by the 
radius of the target two-sphere (here taken to be $1$) and $\omega_2=\mu$
is determined by the ``meson'' mass of the model. In fact, one could prove
this, at least for $Q$ in an infinite (but unknown) subset of $\Z$,
by an obvious modification of the methods of Lin and Yang \cite{linyan}, who
considered the functional $F_0$ with $\mu=0$. 
It is well understood that solitons cannot spin with $\omega>\mu$ because
they become unstable to radiation of mesons. We will see evidence, however,
that if $\mu>1$, solitons generically lose stability and collapse {\em before}
$\omega$ reaches $\mu$, due to the extra nonlinear velocity dependence
generated by the Skyrme term. This is not specific to the model considered
here, and should be generic for Skyrme-type models.

We are able to
get significant analytic insight into the behaviour of isospinning hopfions
by developing a time-dependent extension of the elastic rod model of
hopfions introduced recently by two of us, in collaboration with
Sutcliffe \cite{harspesut}. This extended model, which is of considerable
geometric interest in its own right, predicts a universal
energy-frequency relationship, in rather good agreement with our main 
numerical results. We also test the rod model by comparing its predictions for 
infinite isospinning Hopf strings (i.e.\ solutions which are equivariant
with respect to translations in a fixed direction) with numerics. Again,
we find rather good agreement. 

The rest of this paper is structured as follows. In section \ref{sec:red}
we reduce the problem of finding isospinning hopfions to an energy minimization
problem. In section \ref{sec:ela} we develop an elastic rod model for 
time-dependent Faddeev-Skyrme fields and extract phenomenolgical predictions
on isospinning hopfions and Hopf strings. In section \ref{sec:num}
we present our numerical results, which are briefly summarized
in section \ref{sec:con}.

\section{Reduction to a static variational problem}
\label{sec:red}
\news

Since isorotation involves only rotational symmetry of the target space $S^2$,
it is convenient to consider the Faddeev-Skyrme model on a general oriented
Riemannian manifold $M$. This allows one to treat in unified fashion
the case of principal interest, $M=\R^3$, and the cases of soliton
chains or strings, $M=\R^2\times S^1$, sheets $M=\R\times T^2$, or
geometrically nontrivial domains (of potential interest for cosmological
applications, for example).
Given a time-dependent field $\phi:\R\times M\ra S^2$, we have at each fixed 
time $t$ a mapping $\phi(t,\cdot):M\ra S^2$ which we shall, in a slight abuse 
of notation, again denote $\phi$, and a time derivative $\dot\phi$,
which is a section of the bundle $\phi^{-1}TS^2$ over $M$. Using these,
we define, at time $t$, the
 kinetic and potential energy functionals to be
\bea \label{T-V}
T&=&\int_M\frac12|\dot\phi|^2+\frac12|\phi^*(\iota_{\dot{\phi}}\Omega)|^2\\
V&=&\int_M\frac12|\d\phi|^2+\frac12|\phi^*\Omega|^2+U(\phi),
\eea
where $\Omega$ is the area form on 
$S^2$, $\phi^*\Omega$ its pullback to $M$, $\iota$ denotes interior product, 
and $U:S^2\ra [0,\infty)$ is a smooth potential function
which we assume attains its minimum value $0$ at some point 
$\psi_\infty\in S^2$, and is invariant under rotations about $\psi_\infty$. 
If $M$ is noncompact, we assume that $\phi(t,x)\ra\psi_\infty$, as 
$x\ra\cd_\infty M$, sufficiently fast for all integrals to converge.

Let $\omega>0$ be a fixed constant. We seek time-periodic solutions
 of period $2\pi/\omega$. By definition, these are
critical points $\phi:S^1_\omega\times M\ra S^2$ of the action functional
\beq
S(\phi)=\int_{S^1_\omega}(T-V)
\eeq
where $S^1_\omega=\R/(2\pi/\omega) \Z$ is the circle of length $2\pi/\omega$.  % DH changed 2\pi\omega to 2\pi/\omega twice 
Denote by $X_\omega$ the completion in $C^1$ of the set of smooth maps
$\phi:S^1_\omega\times M\ra S^2$ of finite action. We define an action of
the group $S^1=\R/2\pi\Z$ on $X_\omega$ as follows: 
\beq
([\alpha],\phi)\mapsto\phi_{[\alpha]},\qquad
\phi_{[\alpha]}(t,x)=R(\alpha)\phi(t-\alpha/\omega,x)
\eeq
where $R(\alpha)$ denotes the $SO(3)$ matrix generating rotation through angle
$\alpha$ about the axis $\psi_\infty$. Clearly, $S(\phi_{[\alpha]})=S(\phi)$
for all $([\alpha],\phi)$, since the action is separately invariant under both
time translation and isorotation about $\psi_\infty$. 
Denote by $X_\omega^{S^1}$ the set of fixed points of this action. Then
$\phi\in X_\omega^{S^1}$  if and only if
\beq\label{uif}
\phi(t,x)=R(\omega t)\psi(x)
\eeq
for some map $\psi:M\ra S^2$. We may think of $\psi$ as the
stationary field $\phi$ when viewed in an internally corotating frame.
Since $S^1$ is compact, it follows from the
Principle of Symmetric Criticality \cite{pal} that $\phi\in X_\omega^{S^1}$
is a critical point of $S:X_\omega\ra\R$ if and only if it is a critical 
point of the restricted action $S:X_\omega^{S^1}\ra\R$. Now
\beq
S(R(\omega t)\psi(x))=\frac{2\pi}{\omega}\left\{ % DH changed 2\pi\omega to 2\pi/\omega
\frac12\omega^2\int_M(|\psi_\infty\times\psi|^2+|\d(\psi_\infty\cdot\psi)|^2)
-V(\psi)\right\},
\eeq
so a uniformly isorotating field of the form (\ref{uif}) is a critical point
of $S$ if and only if the static field
$\psi:M\ra S^2$ is a critical point of the
functional
\beq\label{pe}
F_\omega(\psi)=\int_M\left\{\frac12(|\d\psi|^2-\omega^2|\d(\psi_\infty\cdot\psi)|^2)+
\frac12|\psi^*\Omega|^2+(U(\psi)
-\frac12\omega^2|\psi_\infty\times\psi|^2)\right\}.
\eeq
We shall call this functional the {\em pseudoenergy} of $\psi$.
It has a rather interesting and natural form, as we now describe.

The first two terms of (\ref{pe}), taken together, can be interpreted as
the Dirichlet energy of the map $\psi:M\ra S^2$, where $S^2$ is given the
deformed metric
\beq
\ip{X,Y}_\omega=X\cdot Y-\omega^2(\psi_\infty\cdot X)(\psi_\infty\cdot Y)
\eeq
for all $X,Y\in T_\psi S^2$. For $0<\omega<1$ this metric gives $S^2$ the 
geometry of an oblate sphere, squashed along the direction of $\psi_\infty$.
For $\omega> 1$, the metric is singular,
changing from Riemannian to Lorentzian in a strip around the equator
(orthogonal to $\psi_\infty$). Consequently, the pseudoenergy
$F_\omega$ is no longer bounded below for $\omega>\omega_1= 1$ which, as we will
see, has strong phenomenological consequences.

The third term of $F_\omega$ is just the usual Faddeev-Skyrme term (quartic
in spatial 
derivatives). The fourth and fifth terms together can be interpreted as
a deformed potential
\beq
U_\omega(\psi)=U(\psi)-\frac12\omega^2|\psi_\infty\times\psi|^2.
\eeq
Since $U$ attains its minimum at $\psi_\infty$ and is rotationally invariant,
its hessian about $\psi_\infty$ must be $\mu^2\ip{\cdot,\cdot}$ for some
constant $\mu\geq0$, interpreted physically as the mass of mesons in the field
theory. Hence, if $\omega >\omega_2=\mu$, $F_\omega$ is again unbounded below.
A particularly convenient choice for $U$ is 
\beq\label{niceU}
U(\psi)=\frac12\mu^2(1-\psi_3^2)^2.
\eeq
Then $\psi_\infty=(0,0,1)$ and the deformed potential is
\beq
U_\omega(\psi)=\frac12(\mu^2-\omega^2)(1-\psi_3^2)^2.
\eeq
This is the potential we use in all our numerical simulations. 

Since the model is invariant under global rotations of $\phi$ about 
$\psi_\infty$, it has an associated conserved Noether charge, called
{\em isospin}
\beq
J=\int_M\left\{\dot\phi\cdot(\psi_\infty\times\phi)+
\ip{\d(\psi_\infty\cdot\phi),\phi^*(\iota_{\dot\phi}\Omega)}\right\}.
\eeq
For uniformly isorotating fields of the form (\ref{uif}), this equals
\beq
J=\Lambda(\psi)\omega
\eeq
where the moment of inertia is
\beq
\Lambda(\psi)=\int_M\left\{|\psi_\infty\times\psi|^2+|\d(\psi_\infty\cdot\psi)|^2
\right\}.
\eeq
Hence $F_\omega(\psi)=V(\psi)-\frac12\omega^2\Lambda(\psi)$, while the
total energy of the field (\ref{uif}) is $V+T=V(\psi)+J^2/(2\Lambda(\psi))$.
There are two natural variational problems for $\psi$:
\begin{enumerate}
\item For fixed $\omega$, extremize 
$F_\omega(\psi)=V(\psi)-\frac12\omega^2\Lambda(\psi)$;
\item For fixed $J$, extremize
$E_J(\psi)=V(\psi)+J^2/(2\Lambda(\psi))$.
\end{enumerate}
It is clear that these two problems are precisely equivalent: if $\psi$
solves 1, then it solves 2 with $J=\Lambda(\psi)\omega$, and if $\psi'$
solves 2, it solves 1 with $\omega=J/\Lambda(\psi')$. Previous studies 
\cite{batkrusut} of
isorotating\footnote{Actually, \cite{batkrusut} concerns spatially rotating
Skyrmions, but within the axially symmetric ansatz used therein, rotation is 
equivalent to isorotation.}
 solitons have used formulation 2, whereas in this paper we
will mainly use formulation 1.\, This has several advantages. First, it is
directly clear, as we have shown (using the Principle of Symmetric Criticality),
that solutions of 1 correspond via (\ref{uif}) to genuine solutions of the 
field theory. Second,
the Euler-Lagrange equation corresponding to 1 is
a PDE, similar in structure to the static field equation of
the Faddeev-Skyrme model, whereas the equation corresponding to 2 is
a rather more complicated differential-integral equation. Consequently, 
it is a fairly simple matter to adapt
existing numerical techniques, developed for the static
FS model, to deal with problem 1.\, Third, formulation 1 makes it clear that,
in the case $\mu>1$, there is no reason why isospinning solitons should
persist for frequencies $\omega\in (1,\mu)$, since $F_\omega$ is
unbounded below when $\omega>\min\{1,\mu\}$. Hence, we have the possibility
that isospinning hopfions are destabilized by nonlinear velocity terms
in the field equation {\em before} they reach the upper limit $\omega=\mu$.

To conclude this section, we note that a general time dependent solution of
the model $\phi(t,x)=R(\omega t)\psi(t,x)$ conserves both total energy
$T+V$ and isospin $J$, and hence conserves
\beq
T+V-\omega J=F_\omega(\psi)+\frac12\int_M(|\dot\psi|^2+
|\iota_{\dot{\psi}}\Omega|^2).
\eeq
Hence a local minimum of $F_\omega(\psi)$ corresponds to
an orbitally stable solution of the model, since a small perturbation of $\psi$
is trapped close to $\psi$ by conservation of $T+V-\omega J$. So
the solutions found by our numerical method, which can only find minima of 
$F_\omega$, are guaranteed to be orbitally stable. One should note that
conservation of $T+V-\omega J$ does not directly imply that saddle points of
$F_\omega$ are unstable.

\section{An elastic rod model}
\label{sec:ela}
\news

Before embarking on numerical investigation of the variational problems posed in the previous section, it is useful to consider analogous problems in a simple effective model of Faddeev-Skyrme solitons.  A model based on elastic rods was introduced in \cite{harspesut}, which successfully captures the qualitative features of static solitons with small Hopf degree on both $M=\mathbb{R}^3$ \cite{harspesut} and $M=\mathbb{R}^2\times S^1$ \cite{harfos}.  Although originally formulated only in flat 3-space, the model has a geometrically natural extension to any curved Lorentzian 4-manifold representing space-time, allowing one to model
time-dependent solitons.

The degrees of freedom in the elastic rod model consist of a smooth map $x$ from an oriented 2-manifold $\Sigma$ into a 4-manifold $\mathcal{M}$ equipped with a Lorentzian metric $g$, and a section $m$ of the normal bundle $N\Sigma$ of unit length.  The image of $x$ is taken to represent the preimage under $\varphi$ of the point $-\psi_\infty$ in $S^2$ antipodal to $\psi_\infty$, and the section $m$ represents the projection onto $N\Sigma$ of the pull-back under $\varphi$ of a fixed vector in the tangent space at $-\psi_\infty$.  The SO(2) internal symmetry group of the Faddeev-Skyrme model, which rotates the target 2-sphere leaving $\psi_\infty$ fixed, can be identified with the SO(2) structure group of $N\Sigma$.  The latter acts naturally on $m$, since the sphere bundle and the bundle of linear frames for $N\Sigma$ can be identified.  \ignore{There is an obvious action of the diffeomorphism group of $\Sigma$ on $x$ and $m$, and triples $(\Sigma,x,m)$ related by a diffeomorphism are considered equivalent.}

The elastic rod model is defined by an action comprising three terms which describe the effects of stretching, bending, and twisting of the rod.  The stretching term $S_S$ is simply the area of $\Sigma$, as measured using the metric $\gamma$ on $\Sigma$ induced from $g$:
\begin{equation}
S_S(x) := -\int_\Sigma Vol_\gamma.
\end{equation}
This is nothing other than the Nambu-Goto action of string theory.

The bending term $S_B$ is constructed from the second fundamental form $\mathbb{I}$ of $\Sigma$.  The latter is a section of $\mathrm{Sym}^2(T^{\ast}\Sigma)\otimes N\Sigma$ and is defined by the equation,
\begin{equation}
 \mathbb{I}(X,Y) = \pi_N(\nabla^{LC}_X Y) \quad\forall X,Y\in\Gamma(T\Sigma).
\end{equation}
Here $\nabla^{LC}$ denotes the pull-back of the Levi-Civita connexion on $\mathcal{M}$ to $x^\ast (T\mathcal{M})$ and $\pi_N$ denotes the orthogonal projection from $x^\ast (T\mathcal{M})$ to $N\Sigma$.  Since there are two naturally defined quadratic forms on $\mathrm{Sym}^2(T^{\ast}\Sigma)$, there are two ways to build an action from $\mathbb{I}$.  Introducing coordinates $\sigma^a$ on $\Sigma$ and $y^\mu$ on $\mathcal{M}$, the two possible actions built from $\mathbb{I}=\mathbb{I}_{ab}^\mu\d\sigma^a\d\sigma^b\partial_\mu$ are:
\begin{equation}
S_B(x) := \int_\Sigma g_{\mu\nu}\gamma^{ab}\gamma^{cd} \mathbb{I}^\mu_{ab}\mathbb{I}^\nu_{cd} Vol_\gamma \quad\mbox{and} \quad S_B'(x) := \int_\Sigma g_{\mu\nu}\gamma^{ab}\gamma^{cd} \mathbb{I}^\mu_{da}\mathbb{I}^\nu_{bc} Vol_\gamma.
\end{equation}
These correspond to taking either the square of the trace or the trace of the square of the second fundamental form.

While in general one should consider actions involving both terms $S_B$ and $S_B'$, in applications in which $\mathcal{M}$ is flat one may work solely with $S_B$ without any loss of generality.  The reason is that in this situation the equations of motion for both terms agree, as we now prove.  Suppose that the Riemann curvature tensor for $\mathcal{M}$ vanishes.  Then the Riemannian curvature $R$ of $\gamma$ may be expressed in terms of the second fundamental form using Gauss's formula:
\begin{equation}
R_{abcd} = g_{\mu\nu} ( \mathbb{I}^\mu_{ac} \mathbb{I}^\nu_{bd} - \mathbb{I}^\mu_{ad}\mathbb{I}^\nu_{bc} ).
\end{equation}
Contracting with $\gamma^{ac}\gamma^{bd}$ and integrating over $\Sigma$ results in the equation,
\begin{equation}
\int_\Sigma Sc\ Vol_\gamma = S_B-S_B',
\end{equation}
in which $Sc$ denotes the scalar curvature of $\gamma$.  The left hand side of this equation is equal to a linear combination of the Euler characteristic of $\Sigma$ and an integral over the boundary of $\Sigma$, by the Gauss-Bonnet theorem for manifolds with indefinite signature \cite{jee}.  In particular, it is stable to small local perturbations of $x$ and thus has a trivial Euler-Lagrange equation.  Therefore the Euler-Lagrange equations for $S_B$ and $S_B'$ coincide.

The twisting term $S_T$ is built from the derivatives of $m$.  Recall that a connexion $\nabla^N$ can be defined on the normal bundle $N\Sigma$ by orthogonally projecting the Levi-Civita connexion on $x^\ast T\mathcal{M}$:
\begin{equation}
\nabla_X^N m := \pi^N \nabla_X^{LC} m \quad \forall X\in T\Sigma,m\in N\Sigma.
\end{equation}
The twist rate $\theta$ associated to the unit normal vector $m$ is the projection of the covariant derivative of $m$ to the subspace of $N\Sigma$ orthogonal to $m$.  More precisely, $\theta$ is a section of $T^\ast\Sigma$ defined by the equation
\begin{equation}
 \theta(X)Vol_\gamma(Y,Z) = Vol_g(\nabla_X m, m, Y, Z) \quad\forall X,Y,Z\in \Gamma(T\Sigma).
\end{equation}
The twisting term is then
\begin{eqnarray}
 S_T(x,m) := \int_\Sigma \gamma^{ab}\theta_a\theta_b Vol_\gamma = \int_\Sigma\theta\wedge\ast_\gamma\theta.
\end{eqnarray}
Equivalently, the twisting term may defined to be the $L^2$ norm of $\nabla^N m$.

The complete action for the elastic rod model takes the form
\begin{equation}
S(x,m) = A S_S(x) + B S_B(x) + C S_T(x,m),
\end{equation}
with $A,B,C$ positive real parameters.  Of these, two correspond to choices of units while only the dimensionless ratio $B/C$ is non-trivial.

\subsection{Isospinning rods}

From now on the target manifold $\mathcal{M}$ is taken to be $\R^3\times S^1_\omega$ equipped with its standard Lorentzian metric, and $\Sigma$ is taken to be
compact.  Let $Y_\omega$ denote the set of $C^2$ pairs $(x,m)$
consisting of an immersion $x:\Sigma\ra\mathcal{M}$ and a section
$m$ of $N\Sigma\subset x^{-1}T\mathcal{M}$.  For each $m$ denote by $m'$ the unique section of $N\Sigma$ such that $m,m'$ is a positively oriented orthonormal basis, and let $K_{[\alpha]}$ denote the action of translation along $S^1_\omega$ by $\alpha/\omega$ on $\R^3\times S^1_\omega$.  The action of $S^1$ on $Y_\omega$ corresponding to the action on $X_\omega$ is given by
\begin{equation}
(x,m,\alpha) \mapsto (x_{[\alpha]},m_{[\alpha]}),\quad x_{[\alpha]} := K_{[\alpha]}\circ x,\quad m_{[\alpha]} := m\cos\alpha + m'\sin\alpha.
\end{equation}

The set of fixed points of this action will be denoted $Y^{S^1}_\omega$, and elements of this set may be described as follows.  Let $\Upsilon$ be a 
compact 1-manifold, let $\mathbf{x}:\Upsilon\to\R^3$, and let $\mathbf{m}(s)$ be a unit vector perpendicular to $\cd_s \mathbf{x}$.  Suppose that a local coordinate $s$ on $\Upsilon$ has been chosen so that $|\cd_s \mathbf{x}|=1$; then the map is said to be arclength parametrised.  Any element of $Y^{S^1}_\omega$ takes the form
\begin{eqnarray}
\Sigma &=& S^1_\omega\times\Upsilon \\
x(t,s) &=& (t, \mathbf{x}(s)) \\
m(t,s) &=& \cos\omega t\,\mathbf{m}(s)+\sin\omega t\,\mathbf{t}(s)\times\mathbf{m}(s)
\end{eqnarray}
for some $\Upsilon,\mathbf{x},\mathbf{m}$.

Recall that in the standard Frenet formalism for curves the unit tangent vector is defined to be $\mathbf{t}(s):=\cd_s\mathbf{x}(s)$, and the curvature is $\kappa(s):=|\cd_s \mathbf{t}|$.  The action for an element of $Y_\omega^{S^1}$ is $S=-(2\pi/\omega)F_\omega$, where
\begin{equation}
\label{isospinning Lagrangian}
 F_\omega(\Upsilon,\mathbf{x},\mathbf{m}) := \int_{\Upsilon} \left( A-C\omega^2 + B\kappa^2 + C(\mathbf{t}\cdot\cd_s\mathbf{m}\times\mathbf{m})^2\right)\d s.
\end{equation}
The critical points of the restriction of $F_\omega$ to $Y_\omega^{S^1}$ are critical points of $S$ within $Y_\omega$ by the Principle of Symmetric Criticality \cite{pal}.

As in the Faddeev-Skyrme model, the pseudoenergy for the rod model is bounded from below only for a range of values of $\omega$.  Thus $F_\omega$ can only reasonably be expected to have critical points if $\omega^2\leq A/C$, and attention will be restricted to this range.

A crucial property of $F_\omega$, which underpins the subsequent analysis, is that $F_\omega$ is equivalent to $F_0$ up to a rescaling of lengths and energies:
\begin{equation}
\label{scaling identity}
F_\omega(\Upsilon,\mathbf{x},\mathbf{m}) = \sqrt{1-\omega^2C/A}\, F_0\left(\Upsilon,\sqrt{1-\omega^2C/A}\,\mathbf{x},\mathbf{m}\right).
\end{equation}
This means that properties of $F_\omega$ and its critical points can be gleaned directly from properties of $F_0$ with no extra effort, in contrast with the full Faddeev-Skyrme model.

Within the elastic rod model there is a conserved energy $E$ and a conserved isospin $J=-2C\int_\Upsilon\ast\theta$ which are the Noether charges associated with the symmetries of time translation and rotation of $\mathbf{m}$.  For elements of $Y_\omega^{S^1}$ these take the form $E_J=V+J^2/2\Lambda$ and $J=\omega\Lambda$, with
\begin{eqnarray}
V &=& \int_{\Upsilon} \left( A + B\kappa^2 + C(\mathbf{t}\cdot\cd_s\mathbf{m}\times\mathbf{m})^2\right)\d s\quad\mbox{and} \\
\Lambda &=& \int_{\Upsilon} 2C\d s.
\end{eqnarray}
In particular, $V$ is the energy functional for rods introduced in \cite{harspesut} (to see this, note that if $\mathbf{m}=\sin\alpha(s)\mathbf{n}+\cos\alpha(s)\mathbf{b}$ then $\mathbf{t}\cdot\cd_s\mathbf{m}\times\mathbf{m}=\cd_s\alpha-\tau$ with $\tau$ denoting the torsion).

Previous studies \cite{harspesut, harfos} have found that the static rod model defined by $V$ provides a useful approximation to Skyrme-Faddeev solitons if a value of approximately 0.85 is chosen for the dimensionless ratio $C/B$.  The same choice should be equally useful in the Faddeev-Skyrme model with $U\neq0$, since qualitative properties of solitons seem unaffected by the introduction of $U$ \cite{fos}.  In practice it is necessary to impose constraints on elastic rods in order to prevent self-intersections, but these can safely be ignored for the purposes of the present discussion.

The static rod model successfully captures qualitative features of all low-charge Faddeev-Skyrme solitons, with a minor exception at charge 4.  The problem at charge 4 is that the minimal energy soliton is an $\mathcal{A}_{2,2}$ knot consisting of two superposed charge 2 solitons, and the rod model is unable to describe such a configuration.  The charge 4 $\mathcal{A}_{2,2}$ soliton is the only known energy-minimising soliton which is a superposition of lower charge solitons, so this shortcoming of the rod model is very mild.  Thus we expect that useful insight into the variational problems 1 and 2 posed in section 
\ref{sec:red} can be inferred by studying the corresponding problems for rods.

\subsection{Infinite strings}

Before considering solitons on $\R^3$ within the rod model, we will first discuss a simpler situtation of solitons with cylindrical symmetry.  Let $P>0$ and let us replace physical space $M=\R^3$ by $M=\R^2\times S^1_{2\pi/P}$.  The ansatz,
\begin{equation}
\Upsilon = S^1_{2\pi/P},\quad \mathbf{x}(s) = (0,0,s),\quad \mathbf{m}(s) = (\cos(2\pi s/P),\sin(2\pi s/P), 0),
\end{equation}
describes the unique fixed point of a certain action of $S^1\times\mathrm{SO}(2)$ on $Y_\omega^{S^1}$.  It is a critical point of the pseudoenergy at which the potential energy and moment of inertia take the values
\begin{eqnarray}
\label{straight rod V}
V &=& AP + \frac{(2\pi)^2 C}{P}\quad \mbox{and} \\
\label{straight rod Lambda}
\Lambda &=& 2 P C.
\end{eqnarray}

Consider an extension of variational problem 2, in which $E_J$ is extremized with respect to variations in $P$, $\mathbf{x}$ and $\mathbf{m}$ while $J$ is held fixed.  For each $J$ the expression $E_J=V+J^2/2\Lambda$ has a unique minimum amongst $S^1\times\mathrm{SO}(2)$-symmetric configurations.  The values $E_0(J)$, $P_0(J)$ and $\omega_0(J)$ of $E$, $P$ and $\omega$ at the minimum are easily obtained from equations \eqref{straight rod V} and \eqref{straight rod Lambda} and the expression $\omega=J/\Lambda$:
\begin{eqnarray}
E_0(J) &=& \sqrt{ (4\pi)^2AC + \frac{A}{C}J^2 } \\
P_0(J) &=& 2\pi\sqrt{\frac{C}{A} + \frac{J^2}{(4\pi)^2AC}}\\
\omega_0(J) &=& \frac{\frac{A}{C}J}{\sqrt{ (4\pi)^2AC + \frac{A}{C}J^2 }}.
\end{eqnarray}
These equations show that both $E_0(J)$ and $P_0(J)$ grow linearly with $J$ for large values of $J$, and that $\omega$ is approximately constant for large values of $J$.  They also imply that the dimensionless quantity $E_0P_0^{2}\d\omega_0/\d J$ evaluates to $4\pi^2$ at $J=0$.  These qualitative and quantitative predictions will now be tested in the Faddeev-Skyrme model.

Imposing $S^1\times \mathrm{SO}(2)$ symmetry in the Faddeev-Skyrme model results in the following ansatz for $\psi$ in terms of a profile function $f(r)$:
\begin{equation}
\label{vortex ansatz}
\psi = \left( \sin f(r)\cos(\theta+2\pi z/P), \sin f(r)\sin(\theta+2\pi z/P), \cos f(r) \right).
\end{equation}
The expressions for $V$ and $\Lambda$ are
\begin{eqnarray}
V &=& \pi\int_0^\infty \left\{ (f')^2 + \left( \left(\frac{2\pi}{P}\right)^2+\frac{1}{r^2}\right)\sin^2 f\left(1+(f')^2\right) + \mu^2\sin^2f \right\}r\d r \\
\Lambda &=& 2\pi\int_0^\infty \left( 1+(f')^2 \right)\sin^2f\, r\d r.
\end{eqnarray}
The boundary conditions $f(0)=0$ and $f(r)\to\pi$ as $r\to\infty$ are imposed in order to maintain continuity at the origin and finiteness of $V$ and $\Lambda$.

Variational problem 2 may now be studied using the following numerical scheme.  Let $V(\omega,P)$ and $\Lambda(\omega,P)$ denote the values of $V$ and $\Lambda$ at a critical point of $F_\omega=V-\frac12\omega^2\Lambda$.  The values of the functions $V(\omega,P)$ and $\Lambda(\omega,P)$ at any pair $(\omega,P)$ may be determined by first determining a numerical solution $f$ to the Euler-Lagrange equation associated with $F_\omega$ and then numerically evaluating the integral expressions for $V$ and $\Lambda$.  A minimum $E_0(J)$ of $E=V+\frac{1}{2}\omega^2\Lambda$ with $J=\Lambda\omega$ held fixed can be determined by numerically solving the $J$-preserving gradient flow equation for $E$ in $\omega$ and $P$:
\begin{equation}
\begin{aligned}
\dot P &= -\partial_P E(\omega,P) + M \partial_P J(\omega,P) \\
\dot \omega &= -\partial_\omega E(\omega,P) + M \partial_\omega J(\omega,P) \\
M &= \frac{ \partial_P E \partial_P J + \partial_\omega E \partial_\omega J } { \partial_P J \partial_P J + \partial_\omega J \partial_\omega J }.
\end{aligned}
\end{equation}

\begin{figure}[ht]
\centerline{
 \includegraphics{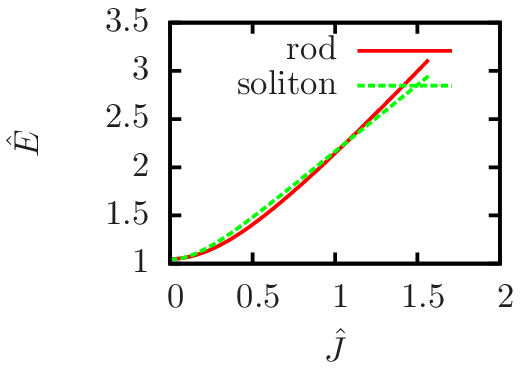}
 \includegraphics{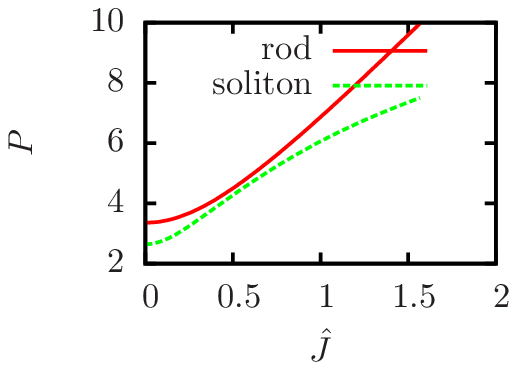}
 \includegraphics{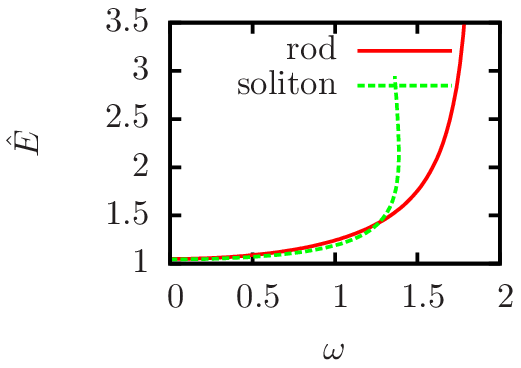}
}
\caption{Infinite strings in the Faddeev-Skyrme model solving variational problem 2.  The graphs show the energy $E$ and period $P$ as functions of the isospin, and also the energy $E$ as a function of the parameter $\omega$.}
\label{fig1}
\end{figure}

The results of this procedure are presented in figure \ref{fig1}.  The numerical results confirm the three qualitative predictions derived in the rod model for large $J$, namely that $E_0(J)$ and $P_0(J)$ grow linearly with $J$, while $\omega_0(J)$ tends to a constant.  From the numerical data we find that
\begin{equation}
\left.E_0P_0^2\frac{\d\omega_0}{\d J}\right|_{J=0} \approx 0.985\times4\pi^2.
\end{equation}
This is within 2\% of the value predicted using the rod model.

To compare the rod and Skyrme-Faddeev models over a range of values of $J$ we must choose values for $A$ and $C$.  We will take
\begin{equation}
(4\pi)^2AC = 1.1,\quad \frac{A}{C} = 3.5.
\end{equation}
The quantity $(4\pi)^2AC$ has the dimension of energy squared, and its value has been chosen to fit the data for the infinite string.  The quantity $A/C$ has the dimension of length to the power of minus two.  Its value has been chosen to give a reasonable fit for both the infinite string and for the solitons discussed below \ignore{(a better match for the infinite string is obtained with the value $A/C=$)}.  The functions $E_0(J)$, $P_0(J)$ and $\omega_0(J)$ in the rod model have been plotted in figure \ref{fig1} alongside those for the Skyrme-Faddeev model.  It can be seen that the curves for $E_0(J)$ are in excellent agreement; the curves for $\omega_0$ and $P_0$ are reasonably close for small $J$, but less close for large values of $J$.  

A final remark concerns the limiting value of $\omega$.  Previously we showed that the Faddeev-Skyrme pseudo-energy is unbounded below for $\omega\geq1$, suggesting that the pseudo-energy does not have critical points for $\omega\geq1$.  Our numerical results for infinite strings confirm that the range of $\omega$ is bounded, however, the maximum value is around 1.4, and exceeds the expected value of 1.  Thus for $1\leq\omega\leq1.4$ the pseudo-energy has critical points, despite being unbounded from below.

\subsection{Solitons}

Now let us consider the effect of isospin on elastic rods in $\R^3$.  Suppose that $(\mathbf{x}_0,\mathbf{m}_0)$ is a critical point of $F_{\omega=0}=V$ and that $V(\mathbf{x}_0,\mathbf{m}_0)=V_0$.  The moment of inertia $\Lambda_0$ for this critical point may be determined by means of a virial theorem: the energy of a rescaled configuration $(\lambda\mathbf{x}_0,\mathbf{m}_0)$ is
\begin{equation}
V(\lambda\mathbf{x}_0,\mathbf{m}_0) = \lambda\int_\Upsilon A \d s_0 + \frac{1}{\lambda} \int_\Upsilon\left( B\kappa_0^2+C(\mathbf{t}_0\cdot\mathbf{m}_0\times\cd_s\mathbf{m}_0)^2\right)\d s_0.
\end{equation}
Since $(\mathbf{x}_0,\mathbf{m}_0)$ is a critical point the derivative of the left hand side with respect to $\lambda$ must evaluate to zero at $\lambda=1$, so the two integrals on the right hand side of the equation must be equal.  It follows that 
\begin{equation}
\Lambda_0 = \frac{C}{A} V_0.
\end{equation}

Since $F_\omega$ is related to $F_0$ by equation \eqref{scaling identity} there exists a corresponding critical point $(\mathbf{x}_0/\sqrt{1-\omega^2 C/A},\mathbf{m}_0)$ of $F_\omega$, and indeed all critical points of $F_\omega$ are so obtained.  The value of $F_\omega$ at the critical point may be deduced from equation \eqref{scaling identity} to be
\begin{equation}
\label{scaling of F}
F_\omega = \sqrt{1-\omega^2C/A}V_0.
\end{equation}
Since $\Lambda$ is proportional to the length of $\mathbf{x}(\Upsilon)$, the value of $\Lambda$ at this critical point is
\begin{equation}
\Lambda = \frac{1}{\sqrt{1-\omega^2C/A}}\frac{C}{A} V_0.
\end{equation}
It follows that $J=\omega\Lambda$ and $E=F_\omega+\omega^2\Lambda$ take the values
\begin{eqnarray}
J &=& \frac{\omega C/A}{\sqrt{1-\omega^2 C/A}} V_0\quad \mbox{and} \\
\label{E(omega)}
E &=& \frac{1}{\sqrt{1-\omega^2C/A}} V_0 \\
\label{scaling of E}
&=& \sqrt{V_0^2 + J^2A/C}
\end{eqnarray}
at the critical point.

The preceding formulae contain useful qualitative information about the solutions to variational problems 1 and 2 in the elastic rod model.  Equation \eqref{scaling of F} implies that the values of $F_\omega$ at its critical points all scale with $\omega$ in the same way.  In particular, the global minimum at $\omega=0$ remains the global minimum of $F_\omega$ for all other values of $\omega$.  As $\omega$ approaches a critical value the moment of inertia, angular momentum, energy and size diverge.

The behaviour of solutions to the second variational problem is similarly uniform: equation \eqref{scaling of E} captures the dependence of the critical values of $E_J$ on $J$.  The global minimum at $J=0$ remains the global minimum for all values of $J$, because the right hand side of \eqref{scaling of E} is monotonic in $V_0$.  At large values of $J$ the energy, moment of inertia and size grow linearly with $J$, while $\omega$ tends to a finite constant.  Similar behaviour should be expected from solitons in the Faddeev-Skyrme model.

\section{Numerical results}
\label{sec:num}
\news

In this section the results of numerical simulations of the Faddeev-Skyrme model will be presented.  Local minima of the pseudo-energy functional $F_\omega(\psi)$ have been found for a range of values of the Hopf degree $Q$ and of $\omega$.  The majority of simulations were carried out with $\mu=2$.  This choice guarantees that $\omega_1<\omega_2$, and thus allows us to investigate the possibility that solitons become unstable to processes other than pion decay.  Other values of $\mu$ were also investigated with Hopf degrees one, two and three.

The numerical algorithm employed was similar to that used in \cite{Jaykka:2011zz}.  A simple 
forward differencing scheme was used on a cubic lattice with a lattice spacing $\Delta x = 0.1$ on a grid typically containing $(240)^3$ latice points.  Well-chosen initial configurations were evolved using the quasi-Newton BFGS method to minimize the pseudoenergy functional $F_\omega(\psi)$.  Simulations were considered to have converged to local minima if the sup norm of the gradient of $F_\omega(\psi)$ was less than $0.01 \Delta x^3$. %edit

Each of our simulations began at $\omega=0$ and proceeded by making small increments in $\omega$.  At each value of $\omega$ the algorithm was allowed to converge to a local minimum of the pseudo-energy, and this local minimum was then used as the initial configuration for the next value of $\omega$.  Initial configurations at $\omega=0$ were created using Sutcliffe's rational map ansatz \cite{Sutcliffe:2007ui}.  This involves identifying physical space $\R^3$ with the three-sphere $S^3\subset\C^2$ using a map 
\beq
(Z_1,Z_0) = \left((x_1+ix_2)\frac{\sin f(r)}{r}, ~~~ \cos f(r) + ix_3 \frac{\sin f(r)}{r} \right),
\eeq
in which $f(r)$ is some monotonic function of $r=\sqrt{x_1^2+x_2^2+x_3^2}$ satisfying the boundary conditions 
$f(0)=\pi, ~f(\infty)=0$.  Initial configurations are given by maps of the form
\beq
W = \frac{\psi_1+i\psi_2}{1+\psi_3} = \frac{Z_1^\alpha Z_0^\beta}{Z_1^a + Z_0^b},
\eeq
with $a,b,\alpha,\beta$ natural numbers.  These maps have Hopf degree $Q=\alpha b + \beta a$, so that an initial configuration with any desired Hopf degree can easily be constructed.  The largest value of $\omega$ that we were able to explore was $\omega=1$, reflecting the fact that the pseudo-energy is unbounded below thereafter.  However, the existence of saddle points of $F_\omega$ with $\omega>1$ is not ruled out.

Below we present location curves and energy density isosurfaces for solitons, and graphs of their energies, both as functions of $\omega$ and as functions of isospin.  As in recent studies \cite{Sutcliffe:2007ui,fos}, the location curves plotted are the tubular preimages of the curve $\psi_3 = -0.90$.

Energy curves obtained from the rod model are plotted alongside those obtained from the numerics, in order to ascertain whether soliton energies follow the expected behaviour.  The rod model curves are obtained from equations \eqref{E(omega)} and \eqref{scaling of E}.  The value of $A/C$ is taken to be 3.5 for all solitons with $\mu=2$, and for each soliton the value of $E_0$ is taken to be the soliton energy at $\omega=0$.  In graphs containing energy curves for several solitons, we choose to plot only one or two curves from the rod model, in order to avoid cluttering the graph.  Graphs are presented using normalised units of energy in which $\hat{E}=E/(16\pi^2\sqrt{2})$ and $\hat{J}=J/(16\pi^2\sqrt{2})$; with this convention Ward's conjectured lower bound on the energy \cite{ward} is $\hat{E}\geq Q^{3/4}$.

In the previous section it was argued using the elastic rod approximation that the sizes and energies of solitons should scale in a uniform way both with $\omega$ and with the isospin $J$.  These predictions are expected to hold for relatively small values of $\omega$ or $J$, but at larger values some deviation from this approximation should be expected.  It is well-known that for most values of $Q$ the Faddeev-Skyrme energy has several local minima with similar energies.  Thus it is plausible that at large values of $J$ or $\omega$ small deviations from the elastic rod approximation can occasionally result in the crossing of energy curves, and hence in a change in the qualitative features of the global minimizer of energy.  The results presented below confirm that the $E(\omega)$ curves follow the predictions of the rod model at small $\omega$. Somewhat surprisingly, the $E(J)$ curves continue to do so even at large $J$. 
 We also observe occasional crossings in both the $E(J)$ and the $E(\omega)$ curves.

\subsection{Degrees one, two and three}

\begin{figure}[ht]
\centerline{
  \includegraphics{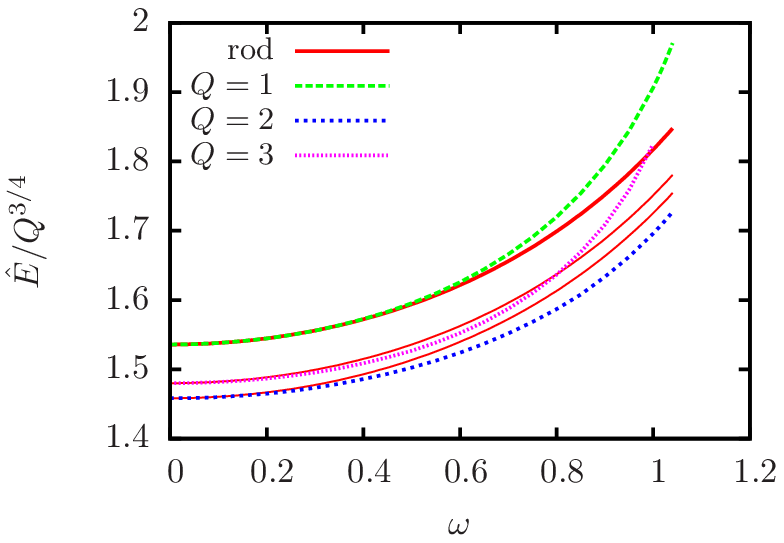}
  \includegraphics{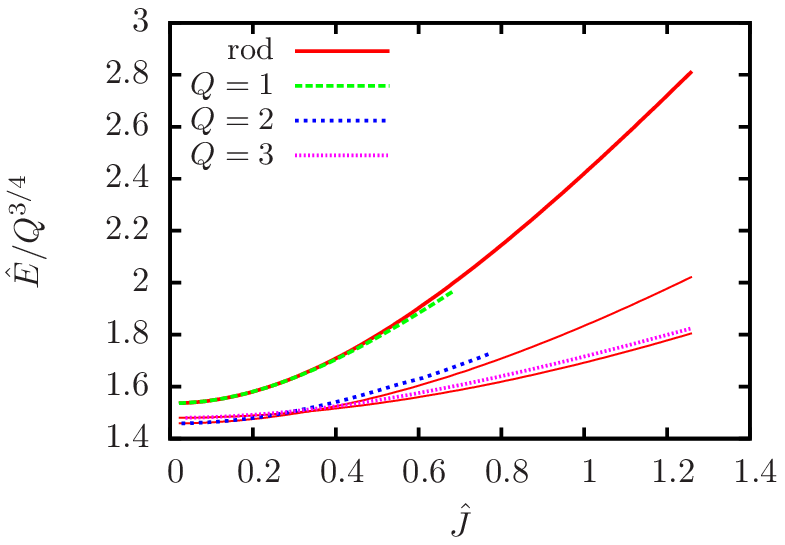}
}
\caption{Energy (in units of $Q^{3/4}$) as a function of angular frequency $\omega$ and as a function of isospin $J$ for solitons with $Q=1,2,3$ and $\mu=2$, and energy curves derived within the rod model.}
\label{fig2}
\end{figure}

\begin{figure}[ht]
   \centering
   \includegraphics[width=5cm]{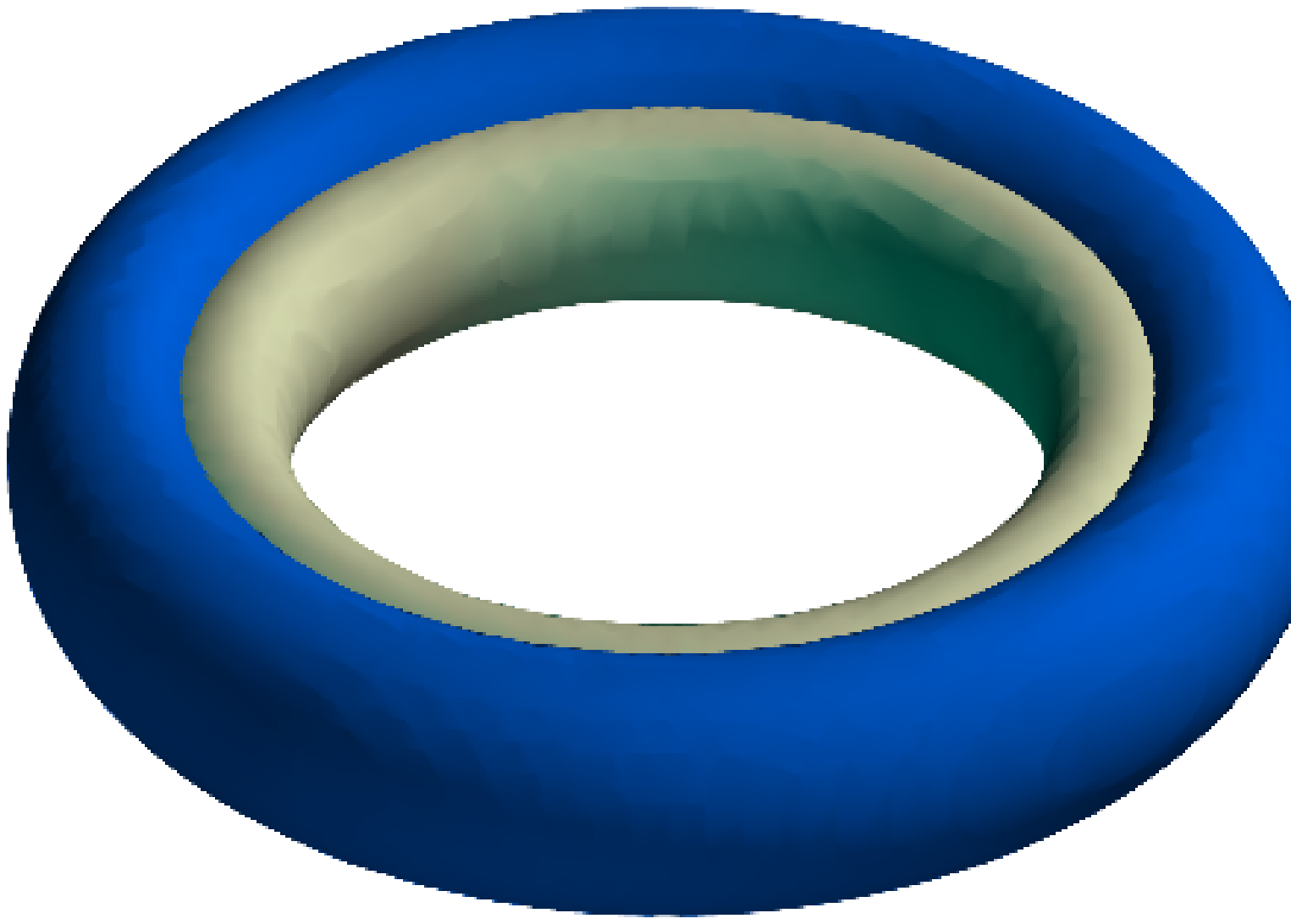} 
   \includegraphics[width=5cm]{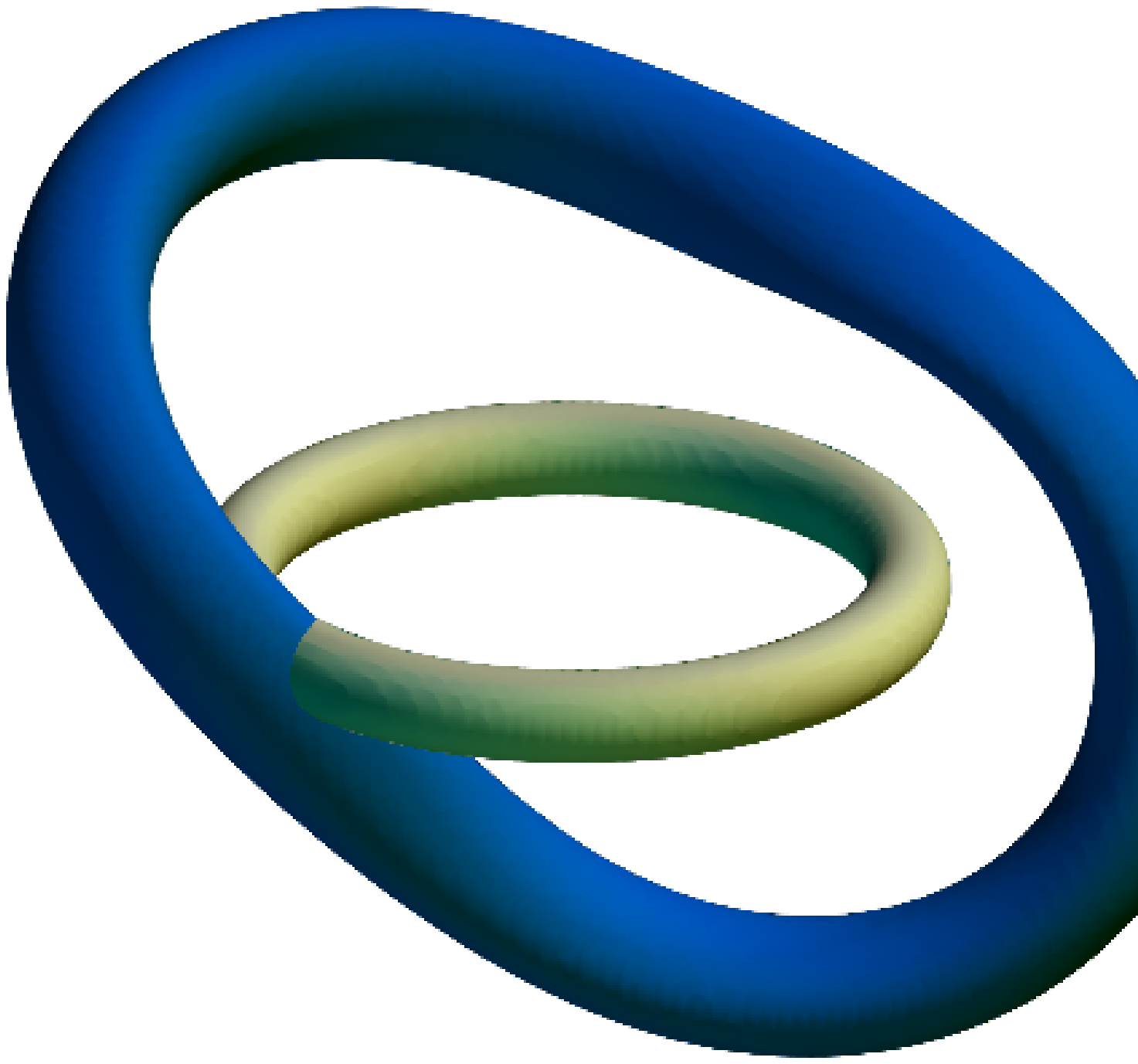} 
   \includegraphics[width=5cm]{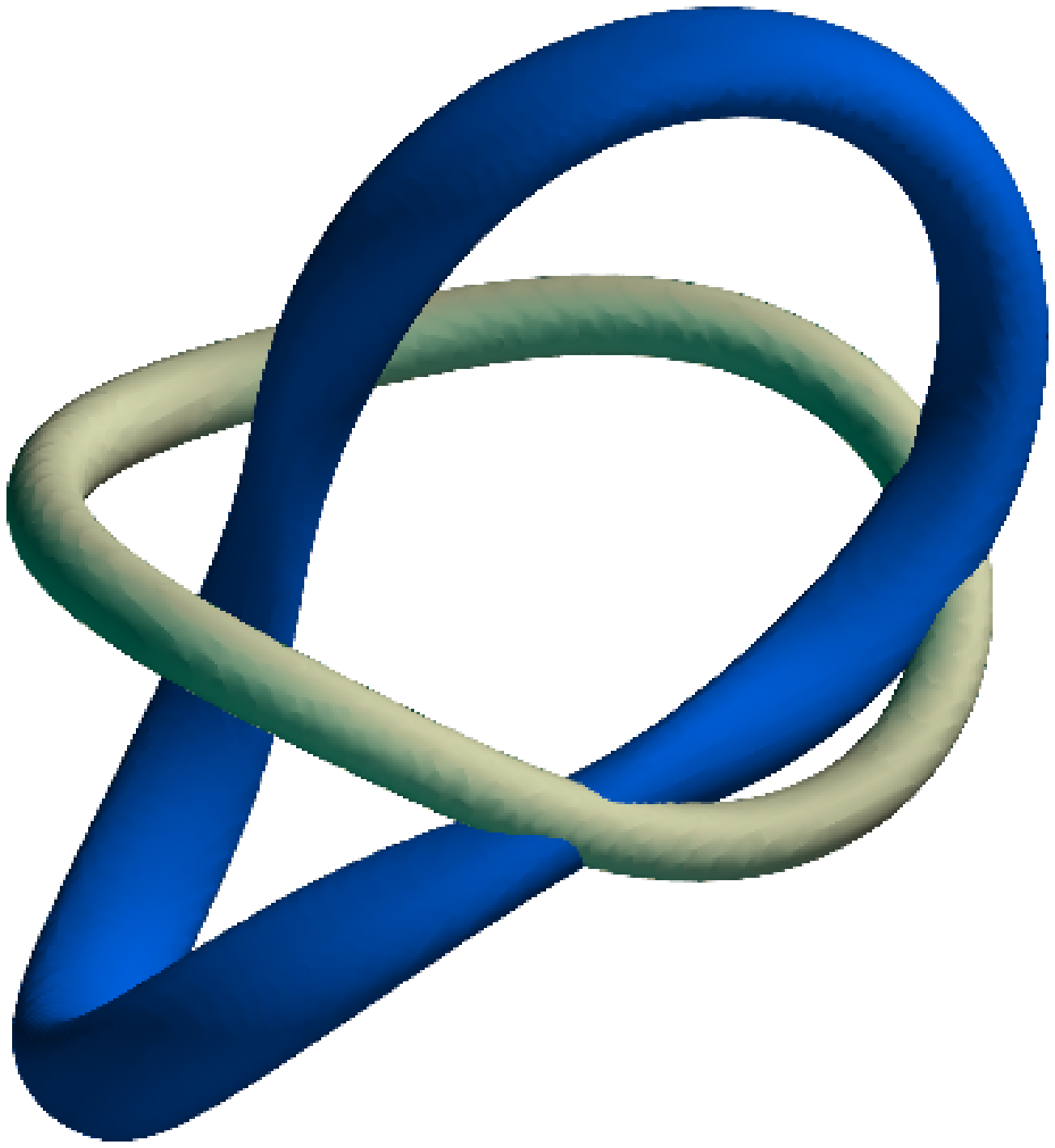}
\centerline{
$1{\cal A}_{1,1}$ \hspace{4.5cm} $2{\cal A}_{2,1}$\hspace{4.5cm}$3{\cal \tilde A}_{3,1}$}
   \caption{Position curves for solitons with charges 
$1\le Q \le 3 $ at $\mu=1$.  The light green and blue position curves correspond to $\omega =0$ and $\omega=1$, respectively.}
   \label{fig3}
\end{figure}

\begin{figure}[ht]
\centerline{
 \includegraphics{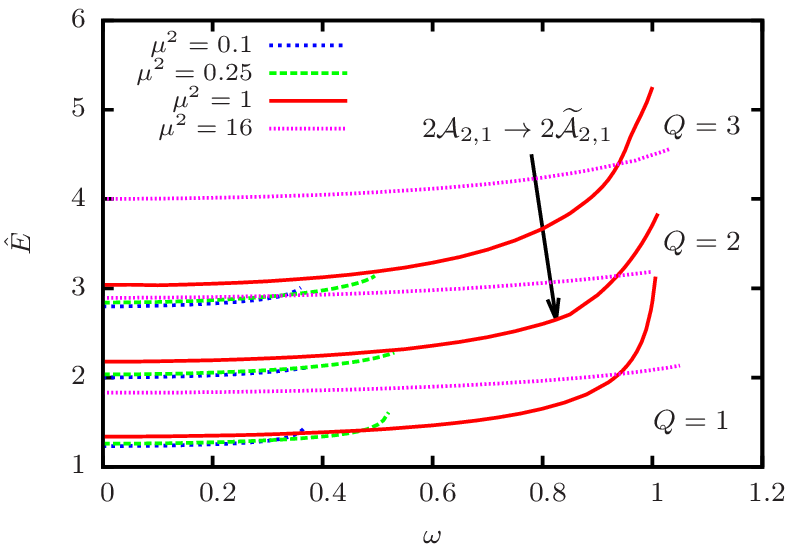}
 \includegraphics{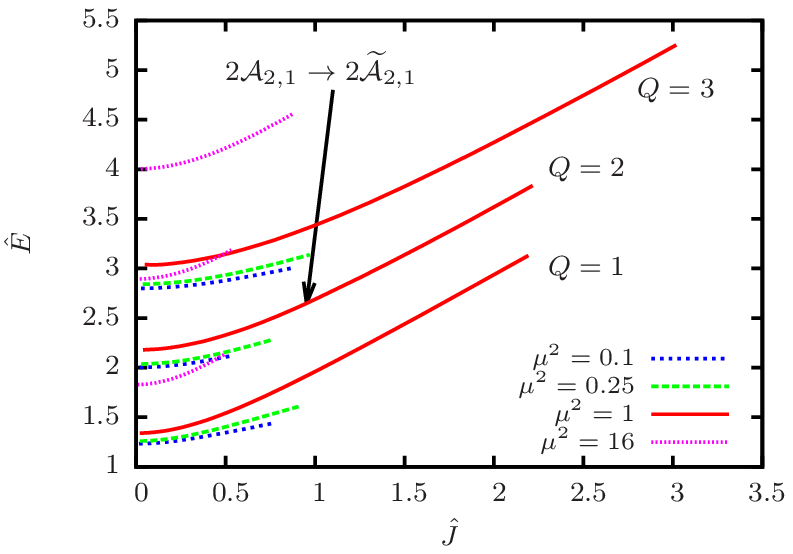}
}
\caption{Energy as a function of angular frequency $\omega$ and as a function of isospin $J$ for $Q=1,2,3$ for a range of values of $\mu$.  The transition $2\mathcal{A}_{2,1}\to\widetilde{\mathcal{A}}_{2,1}$ at $\mu=1$ and $\omega=0.86$, is labelled.}
\label{fig4}
\end{figure}

For Hopf degrees 1, 2 and 3 simulations were carried out with a range of values of $\mu$.  The energy curves for $\mu=2$ are plotted in figure $\ref{fig2}$ alongside the predictions from the rod model.  These clearly show that the soliton energies scale as expected.

In figure \ref{fig4} soliton energy curves for a range of values of $\mu$ are plotted.  Figure \ref{fig3} shows the soliton location curves for $\mu=1$ at $\omega=0$ and $\omega=1$.  For most values of $\omega$ and $\mu$ the preferred configurations are of type ${\cal A}_{1,1}$, $2{\cal A}_{2,1}$ and $3{\cal \widetilde A}_{3,1}$, as in the static case. Here, and in the sequel, we are using the
notation introduced by Sutcliffe to label Hopf soliton shapes \cite{Sutcliffe:2007ui}. 
Briefly, $Q{\cal A}_{n,m}$ denotes an axially symmetric hopfion of charge $Q$,
where the position curve (preimage of $-\psi_\infty$) is a circle, and the
preimage of a regular value close to $-\psi_\infty$ is a disjoint union of
$m$ closed curves, each winding $n$ times around the circle. In fact $Q=nm$,
so including $Q$ in the label is redundant, but convenient. A soliton of
type $Q{\cal \widetilde{A}}_{n,m}$ has the same qualitative form, but with axial
symmetry weakly broken, so the position curve is not exactly circular.
Later we will encounter solitons whose position curves are links of two 
components. These will be denoted $Q{\cal L}_{p,q}^{a,b}$ where the subscripts 
denote the Hopf charges of each component, and the superscripts denote the
extra Hopf degree of each component due to its linking with the other
component. Again, we include $Q$, though this is redundant (since $Q=p+q+a+b$).
We will also encounter hopfions whose position curves are torus knots of type
$(a,b)$ (where $a$ and $b$ denote the windings of the curve around the
$S^1$ factors in $T^2$). 
We denote these $Q{\cal K}_{a,b}$ 

When $Q=2$, $\mu=1$ and $\omega\geq0.86$, our algorithm converges to a buckled $2{\cal \widetilde A}_{2,1}$ rather than an axial $2{\cal A}_{2,1}$ configuration.  This configuration did not appear in the earlier study \cite{acuhalnorshn}, whose attention was restricted to axial symmetry.  The buckled configuration seems to be a local minimum of the pseudo-energy, so should be orbitally stable (as explained in section \ref{sec:red}).  It is likely that an axial $2{\cal A}_{2,1}$ configuration continues to exist above $\omega=0.86$, albeit as a saddle point rather than a minimum of $F_\omega$.  Since our algorithm only finds local minima and not saddle points we are unable to determine whether the $2{\cal \widetilde A}_{2,1}$ has a greater or smaller energy than the $2{\cal A}_{2,1}$ for any given value of $\omega$ or $J$.

When $\mu<1$ our simulations terminate at $\omega=\mu$, as expected.  When $\mu>1$ our algorithm ceases to find any critical points when $\omega=1$ (although in a few cases it is possible to go slightly beyond $\omega=1$).  This is again consistent with our expectations, as the pseudo-energy is not bounded from below when $\omega>1$.  However, the existence of solutions with $\omega>1$ is not ruled out, as they may continue to exist as saddle points of the pseudo-energy.

The borderline case $\mu=1$ is particularly interesting: the graphs of $E(\omega)$ grow rapidly as $\omega$ approaches 1.  The rod model predicted that soliton energies would diverge at some finite value of $\omega$, and our numerical results suggest that for $\mu=1$ this value is very close to 1.

\subsection{Degree four}

\begin{figure}[ht]
   \centering 
   \includegraphics[width=5cm]{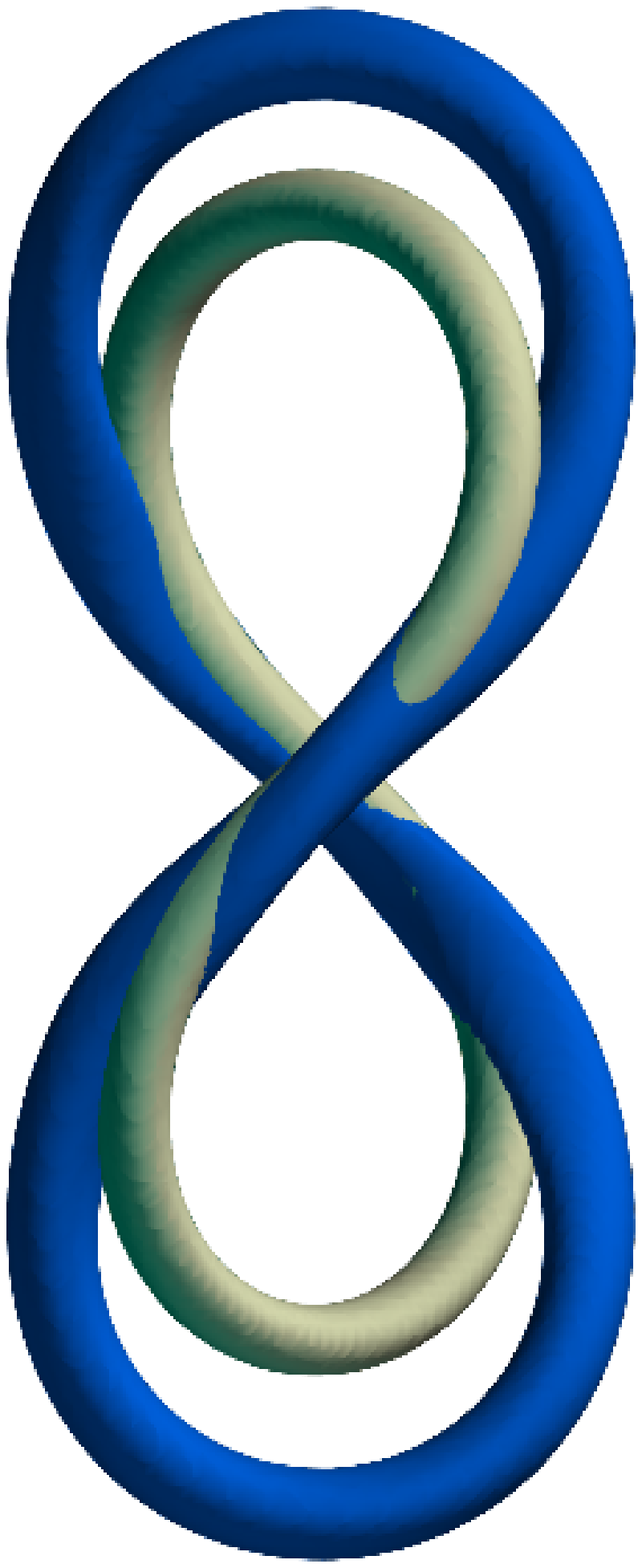}
   \includegraphics[width=5cm]{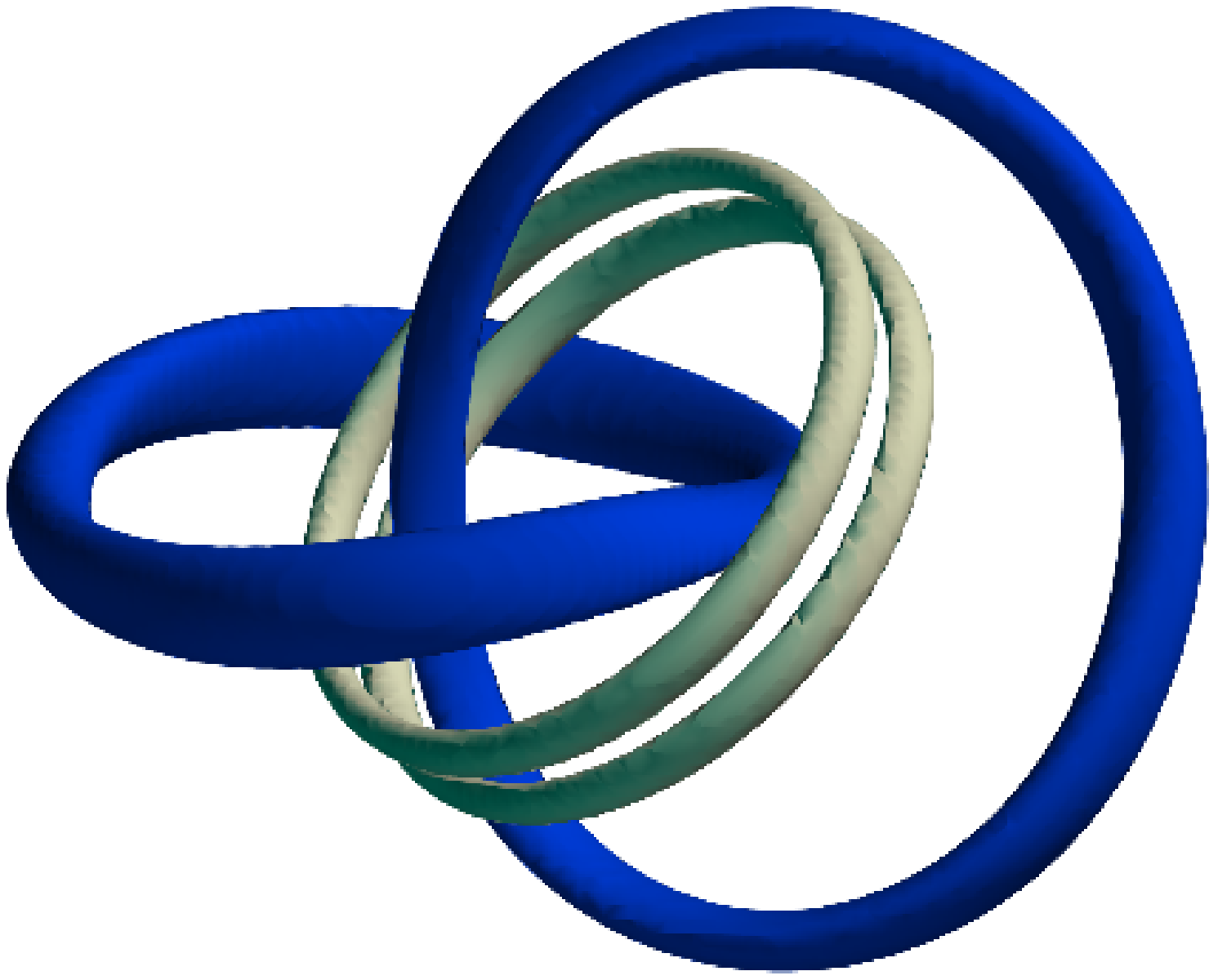}
   \includegraphics[width=5cm]{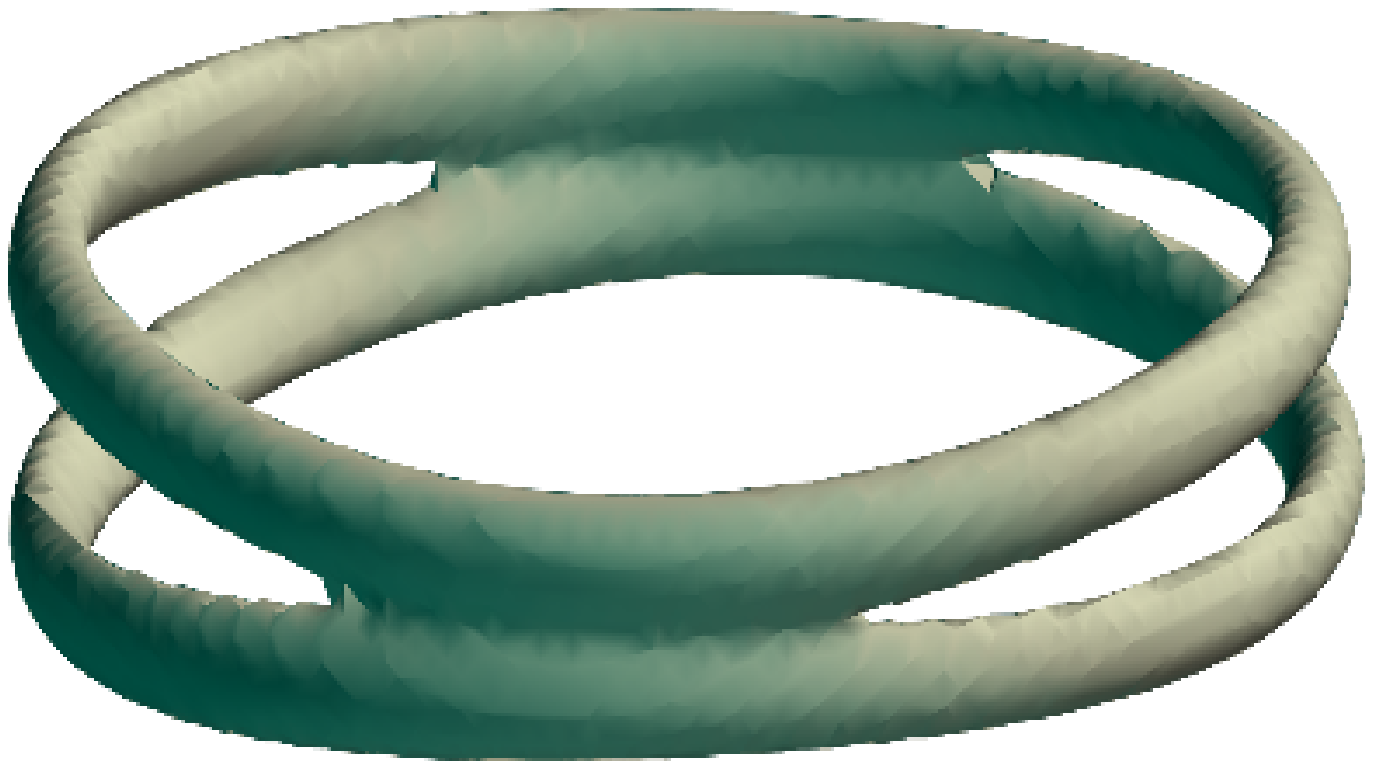}
\centerline{ $4{\cal \tilde A}_{4,1}$ \hspace{3.5cm} $4{\cal A}_{2,2}/4{\cal L}_{1,1}^{1,1}$  \hspace{3.5cm} saddle}
\caption{Position curves for solitons with $Q=4$ and $\mu=2$.  In the left and middle pictures the light green and blue position curves correspond to $\omega =0$ and $\omega=1$ respectively.  The right picture is an intermediate state between $4{\cal A}_{2,2}$ and $4{\cal L}_{1,1}^{1,1}$ at $\omega=0.58$.}
   \label{fig5}
\end{figure}

\begin{figure}[ht]
\centerline{
  \includegraphics{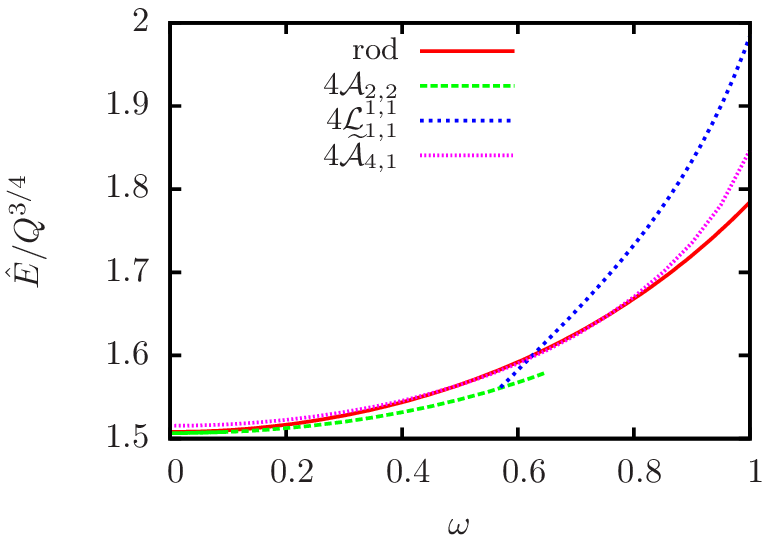}
  \includegraphics{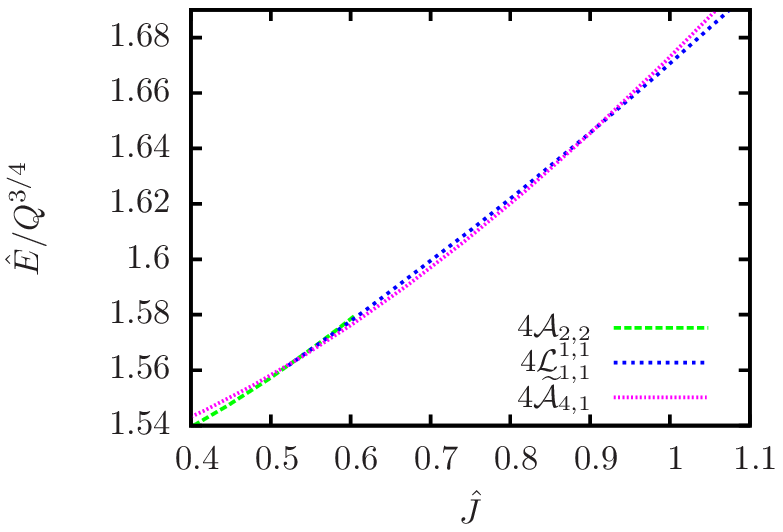}
}
\caption{Energy as function of the parameter $\omega$ and as a function of isospin $J$ for solitons with $Q=4$ and $\mu=2$. In the right panel we only present
data
in the range $0.4\leq\hat{J}\leq 1.1$ to better illustrate the energy 
crossings.}
\label{fig6}
\end{figure}

At degree 4 we have been able to find configurations of types $4 {\cal A}_{2,2}$, $4 {\cal \tilde A}_{4,1}$ and $4{\cal L}_{1,1}^{1,1}$; these are depicted in figure \ref{fig5}.  The $4 {\cal A}_{2,2}$ is axially symmetric and may be thought of as two adjacent $2 {\cal A}_{2,1}$ solitons.  The position curve of the $4 {\cal A}_{2,2}$ in our model consists of two adjacent circles, whereas in the massless Faddeev-Skyrme model it is a single circle.

The graphs of energy as a function of $\omega$ or $J$ are shown in figure \ref{fig6}.  The $4 {\cal \tilde A}_{4,1}$ configuration exists for all values of $\omega$ in the range [0,1], but the $4 {\cal A}_{2,2}$ and $4{\cal L}_{1,1}^{1,1}$ configurations could only be found in the ranges [0,0.65] and [0.57,1] respectively.  At $\omega=0.57$ the $4 {\cal A}_{2,2}$ and $4{\cal L}_{1,1}^{1,1}$ configurations are degenerate in both energy and pseudo-energy, from which we infer that the $4 {\cal A}_{2,2}$ critical point undergoes a bifurcation at this point.  We were also able to find a configuration which seems to be a saddle point of the pseudo-energy at $\omega=0.58$, suggesting that the bifurcation is a pitchfork bifurcation.  This saddle is illustrated in figure \ref{fig5} but its energy has not been plotted.

When $\omega>0.65$ our algorithm is unable to find the $4 {\cal A}_{2,2}$ and instead converges to the $4{\cal L}_{1,1}^{1,1}$.  Since the $4 {\cal A}_{2,2}$ configuration is axially symmetric it is likely to continue to exist as a critical point of the pseudo-energy beyond $\omega=0.65$; the most likely reason for our code's failure to find it is that it is a saddle point rather than a local minimum.

Both the $E(\omega)$ and the $E(J)$ graphs exhibit crossings.  The minimum of the energy at $\hat{J}=0$ is the $4 {\cal A}_{2,2}$ configuration.  All three energy graphs cross in the range $0.5\leq\hat{J}\leq1.0$, and for $\hat{J}\geq1.0$ the energy-minimzer is the $4{\cal L}_{1,1}^{1,1}$ link.  It is likely that the energy-minimizer for fixed $\omega$ is always the $4 {\cal A}_{2,2}$, but for $\omega\geq0.65$ it is difficult to be sure of this without being able to construct the $4 {\cal A}_{2,2}$ configuration.  We note that, despite the presence of crossings, the energy curves generally follow the rod model predictions, the only significant deviant being the $E(\omega)$ curve for the link when $\omega>0.6$.

\subsection{Degrees five to eight}

\begin{figure}[ht]
\centerline{ $5 \widetilde {\cal A}_{5,1}$ \hspace{6cm} $5{\cal L}_{1,2}^{1,1}$ }
\centerline{
   \includegraphics[width=7.5cm]{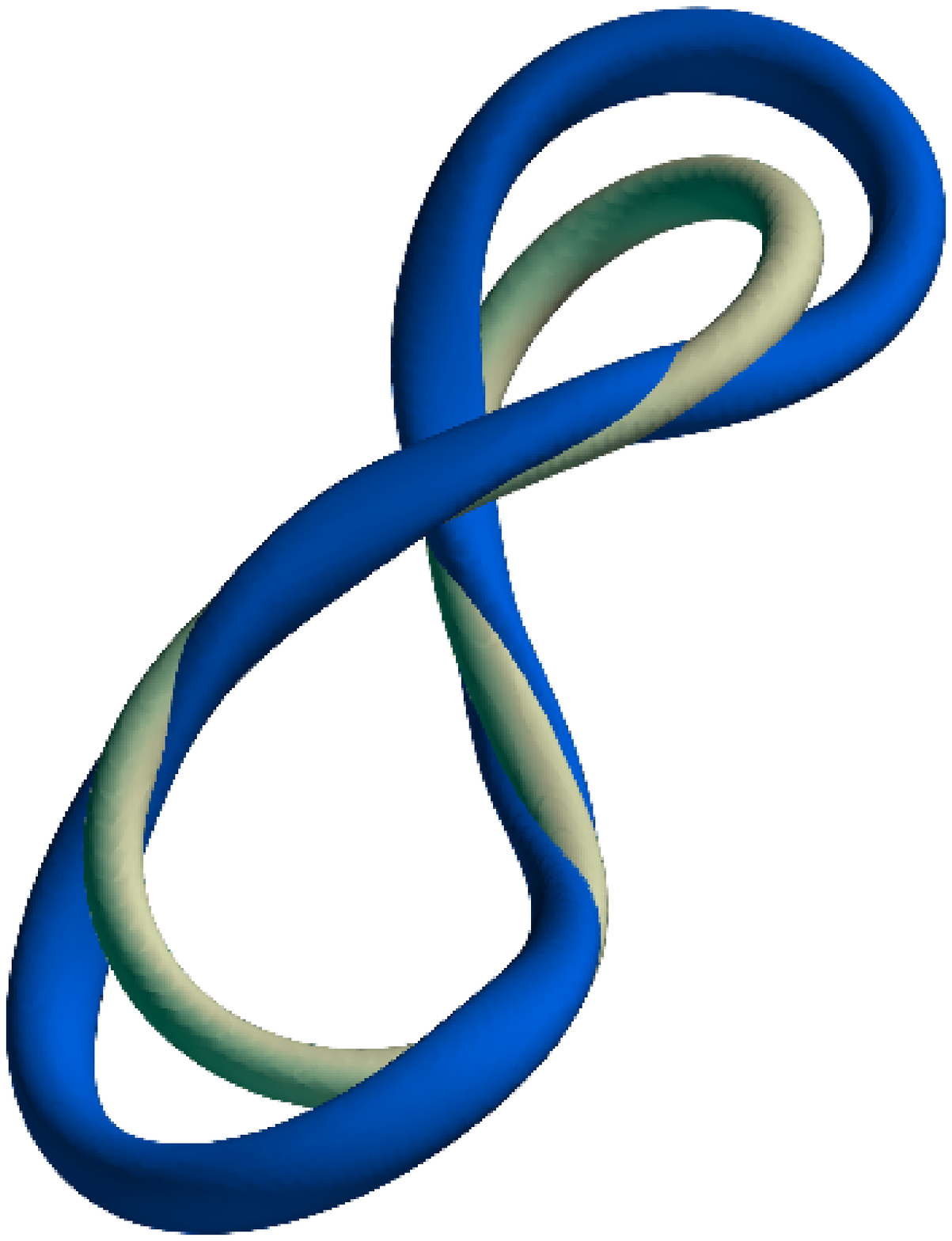} 
   \includegraphics[width=7.5cm]{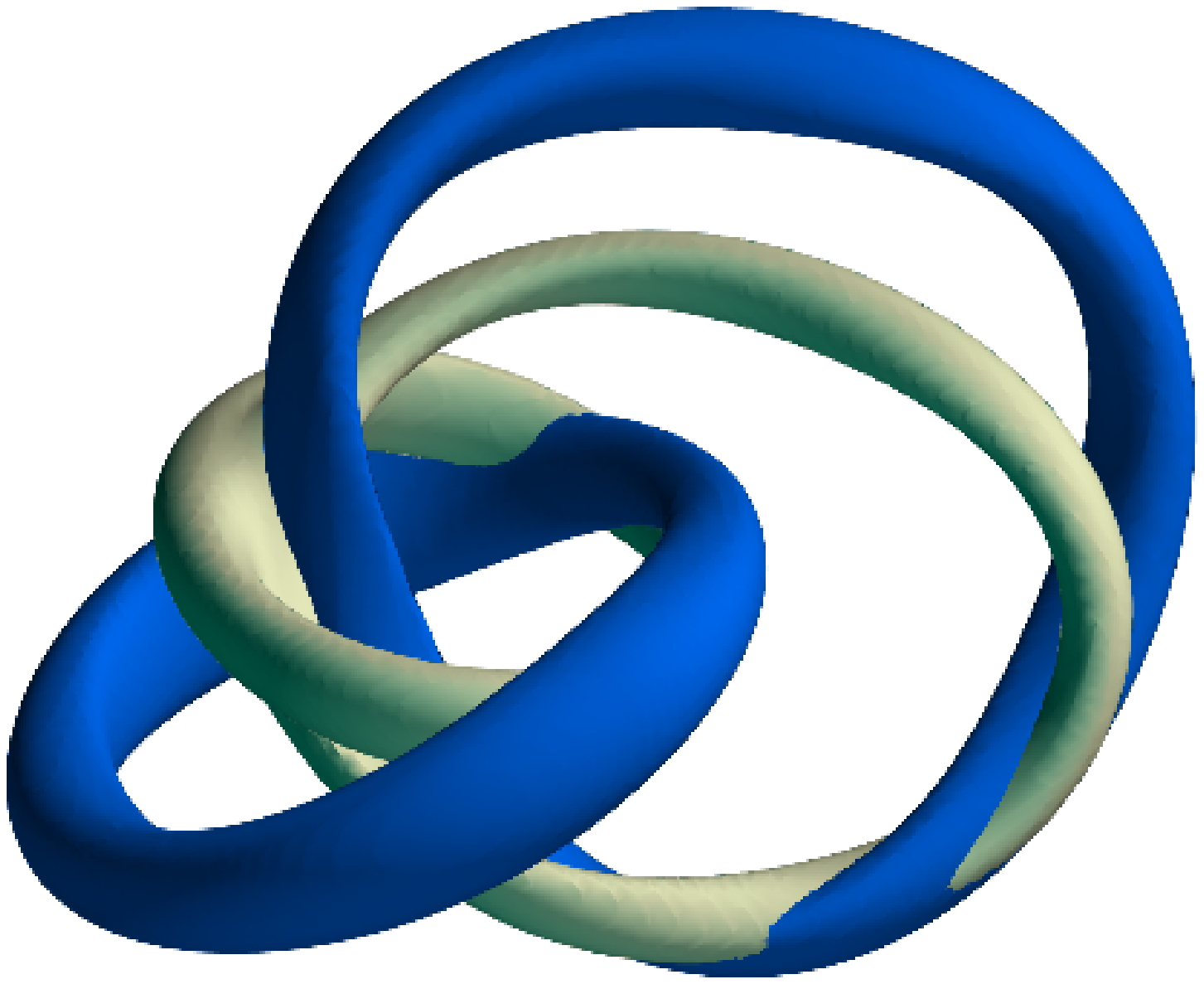}
}
\centerline{
  \includegraphics{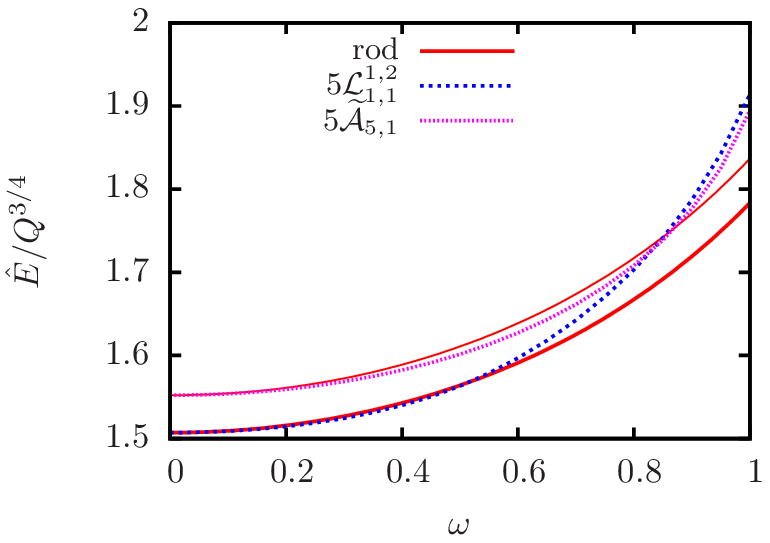}
  \includegraphics{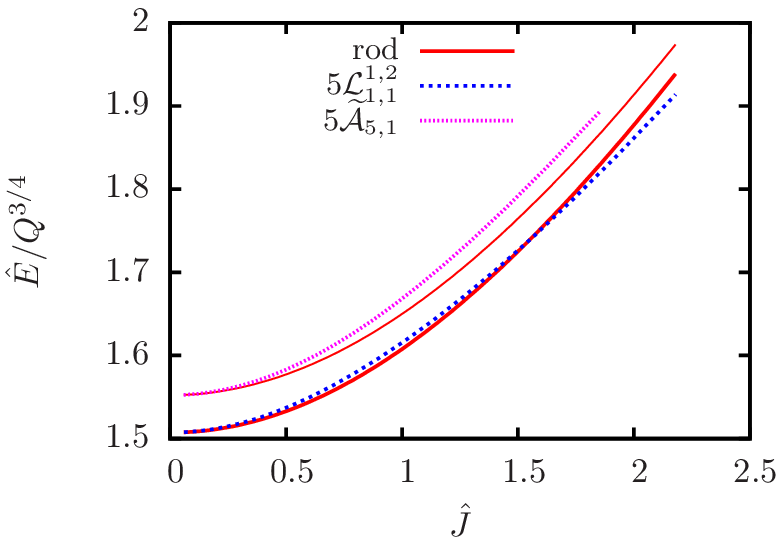}
}
   \caption{Solitons with $Q = 5 $ and $\mu=2$. Top row: position curves for the $5 \widetilde {\cal A}_{5,1}$ 
and  $5{\cal L}_{1,2}^{1,1}$ configurations, with light green and blue position curves corresponding to $\omega =0$ and $\omega=1$ respectively.  Bottom row: energy as function of the parameter $\omega$ and as a function of isospin $J$.}
   \label{fig7}
\end{figure}

\begin{figure}[ht]
\centerline{ $6{\cal L}_{3,1}^{1,1}$ \hspace{5cm} $6 {\cal L}_{2,2}^{1,1}$ }
\centerline{
   \includegraphics[width=7.5cm]{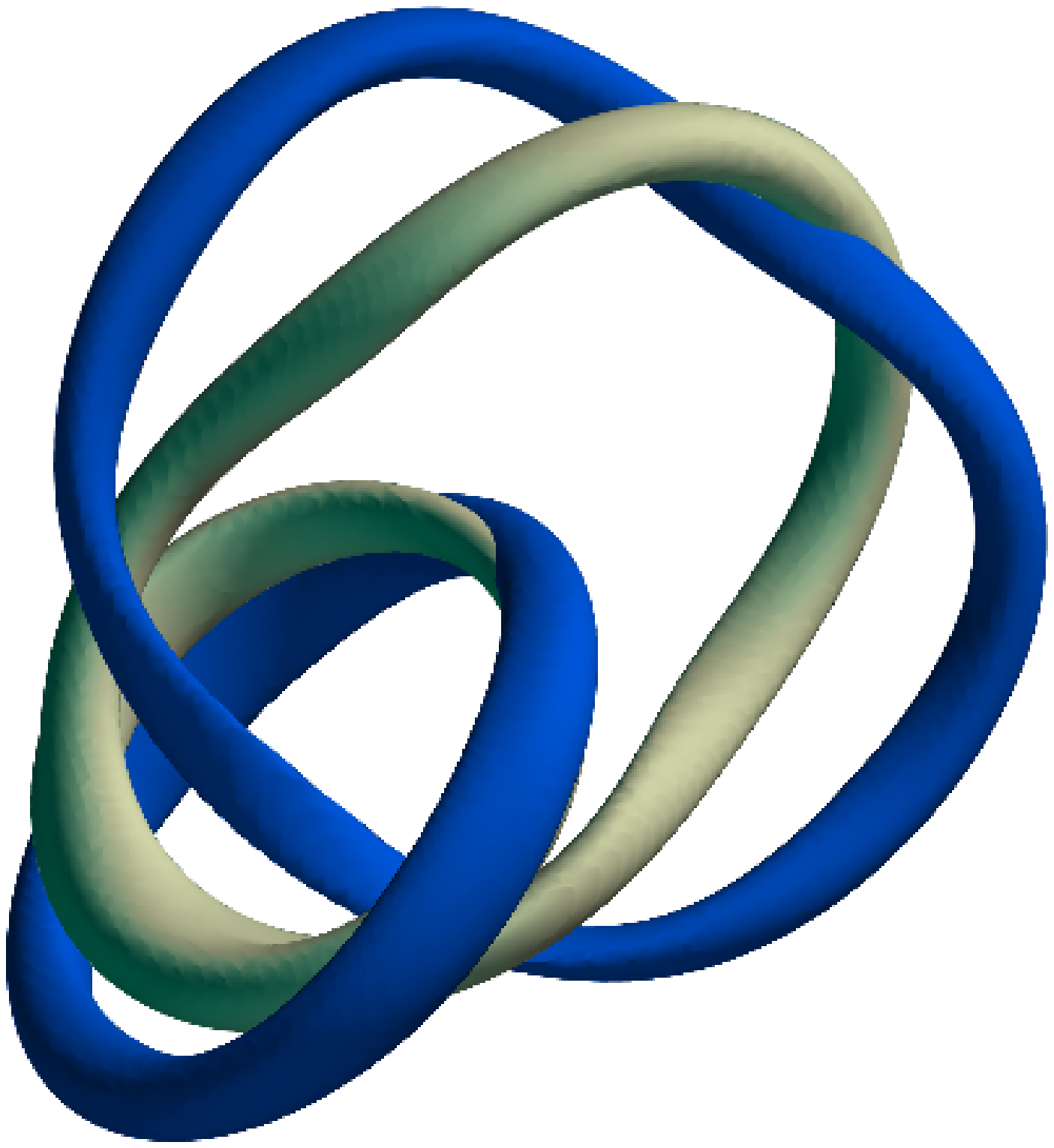}
   \includegraphics[width=7.5cm]{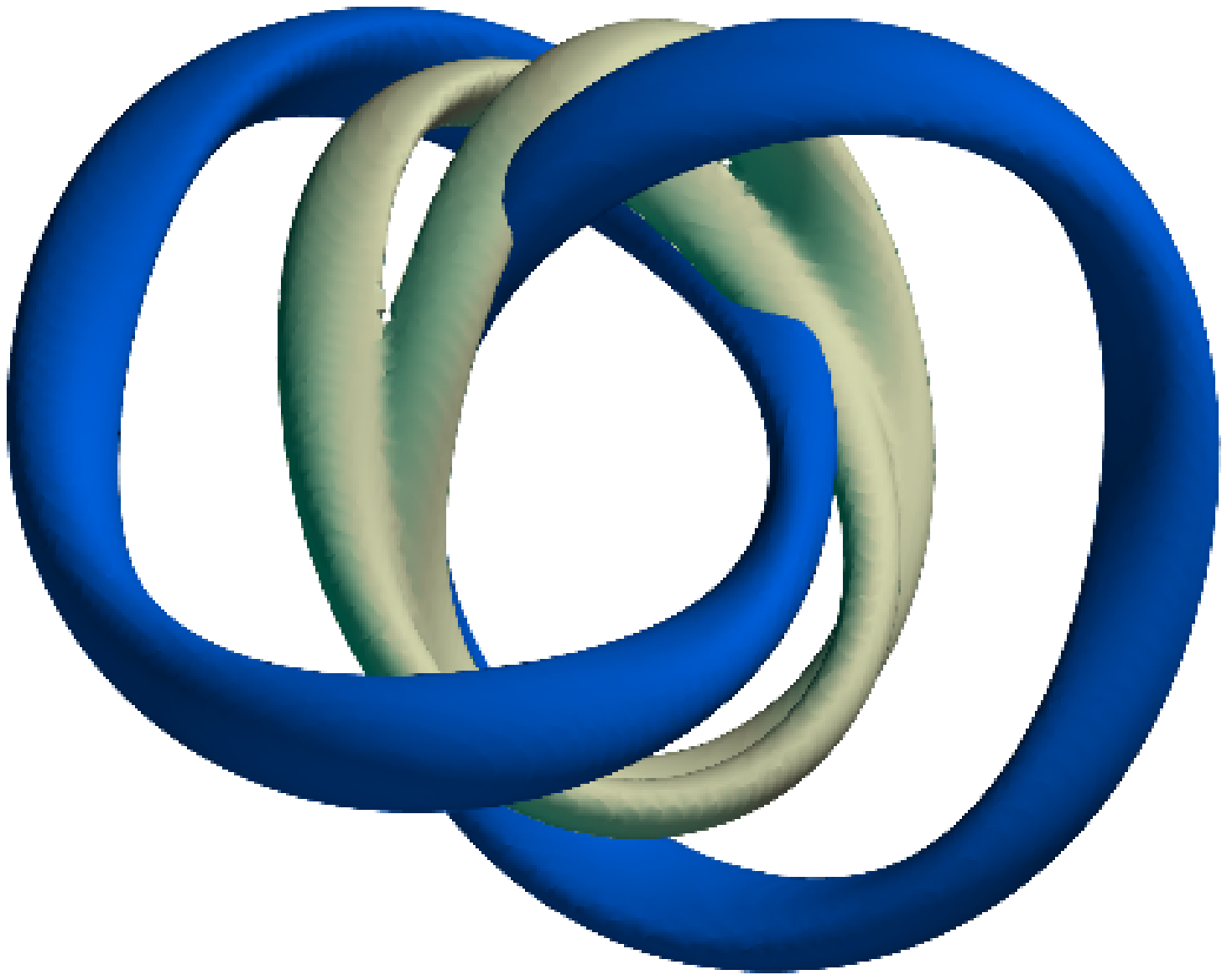}
}
\centerline{
  \includegraphics{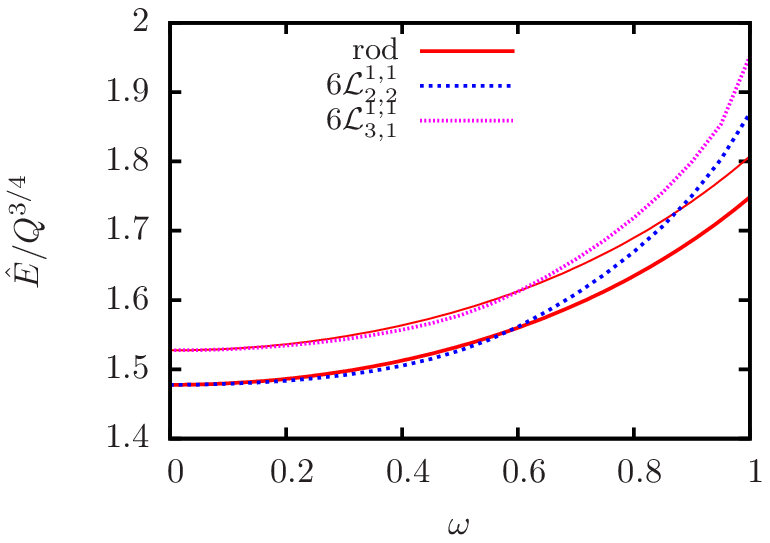}
  \includegraphics{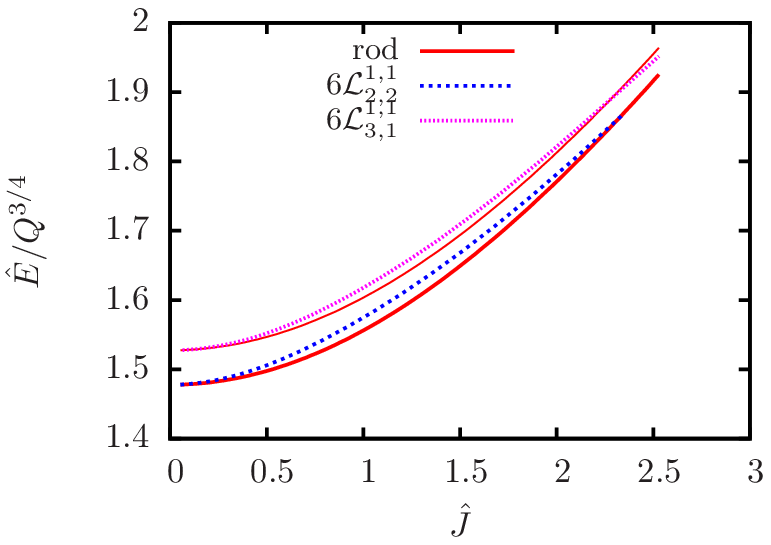}
}
   \caption{Solitons with $Q=6$ and $\mu=2$.  Top row: position curves for the $6 {\cal L}_{2,2}^{1,1}$ 
and  $6{\cal L}_{3,1}^{1,1}$ configurations, with light green and blue position curves corresponding to $\omega =0$ and $\omega=1$ respectively.  Bottom row: energy as function of the parameter $\omega$ and as a function of isospin $J$.}
\label{fig8}
\end{figure}

\begin{figure}[ht]
\centerline{
  \includegraphics[width=5.5cm]{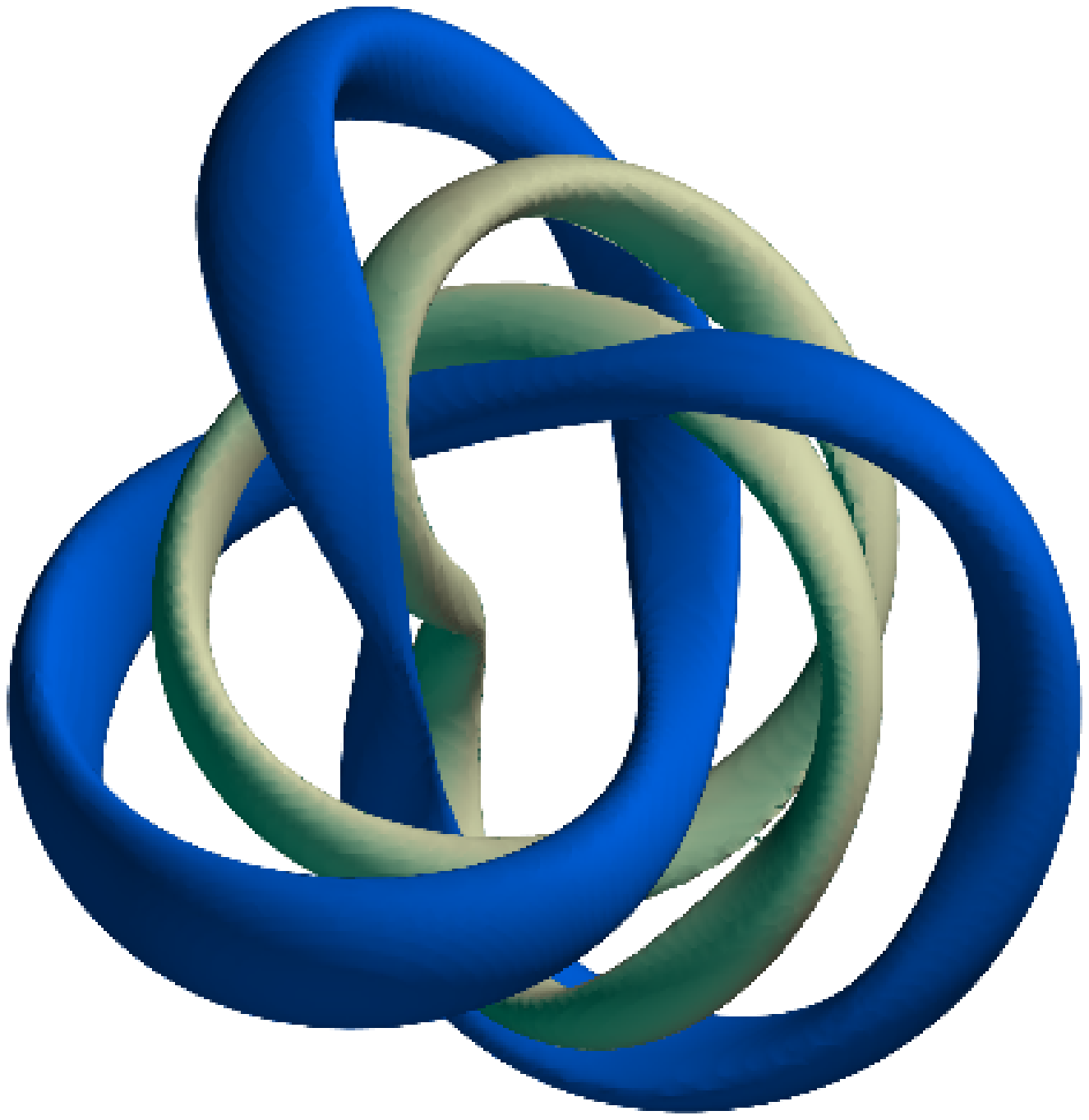} 
  \includegraphics{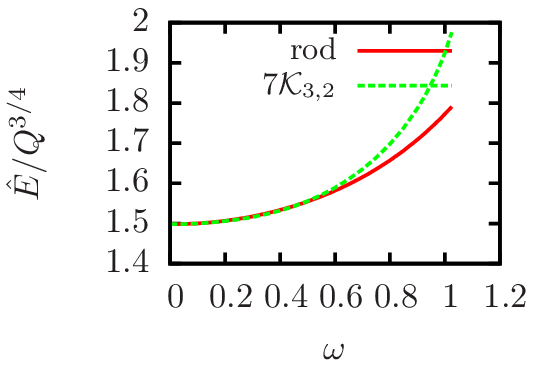}
  \includegraphics{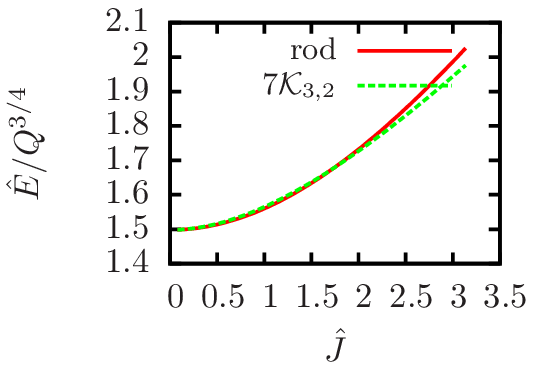}
}
\caption{Solitons with $Q=7$ and $\mu=2$.  The light green and blue curves in the left figure are position curves for solitons at $\omega=0$ and $\omega=1$ respectively.  The two graphs show energy as a function of the parameter $\omega$ and as a function of isospin $J$.}
   \label{fig9}
\end{figure}

\begin{figure}[ht]
\centerline{ $8{\cal K}_{3,2}$ \hspace{4.5cm} $8{\cal L}_{3,3}^{1,1}/8{\cal K}_{3,2}$}
\centerline{
   \includegraphics[width=5cm]{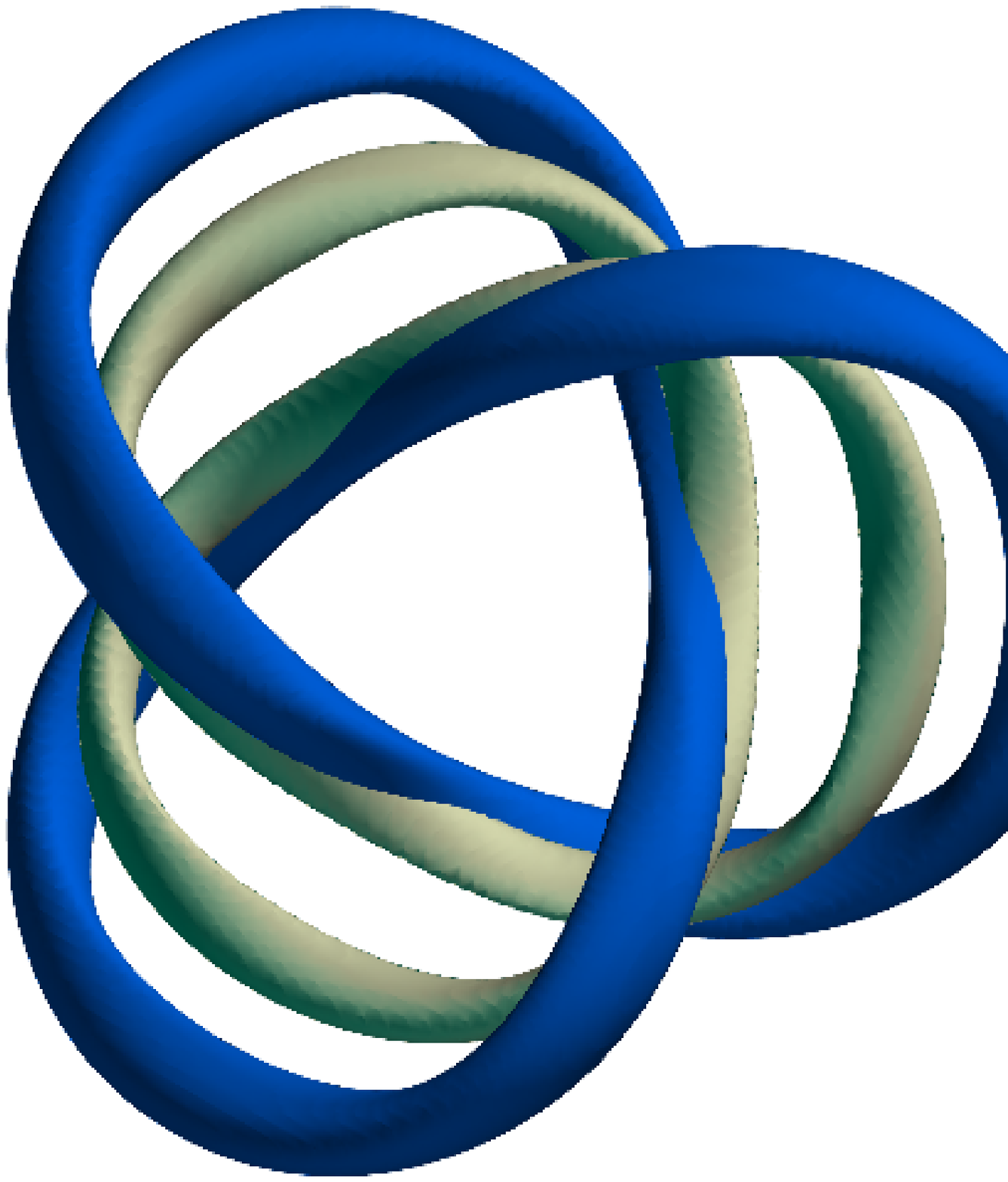} 
   \includegraphics[width=5cm]{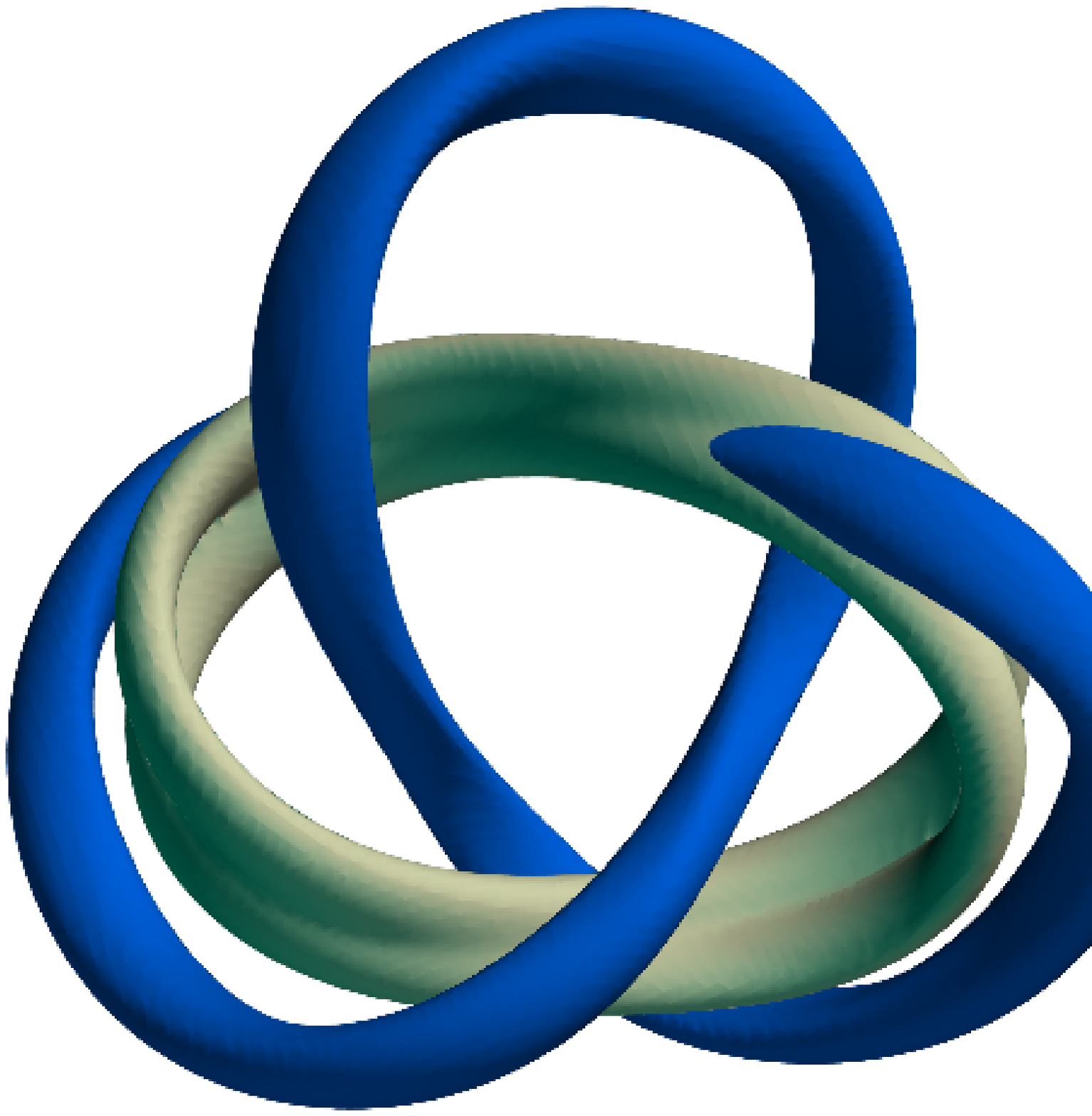}
}
\centerline{
  \includegraphics{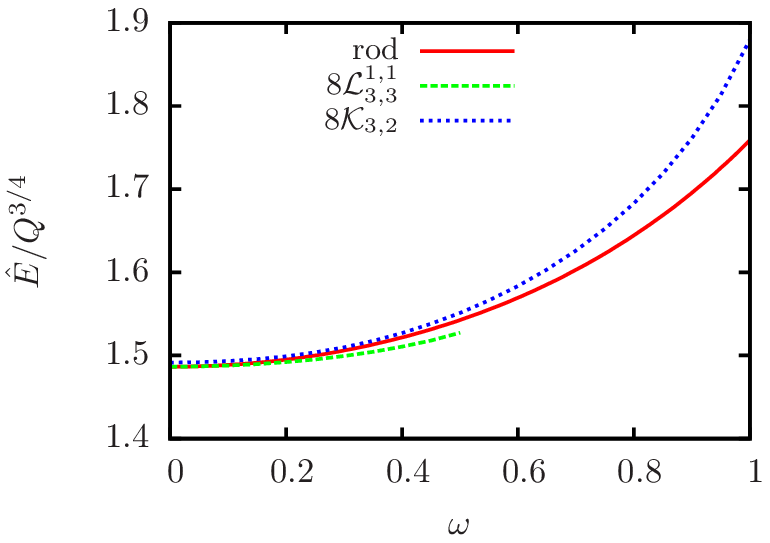}
  \includegraphics{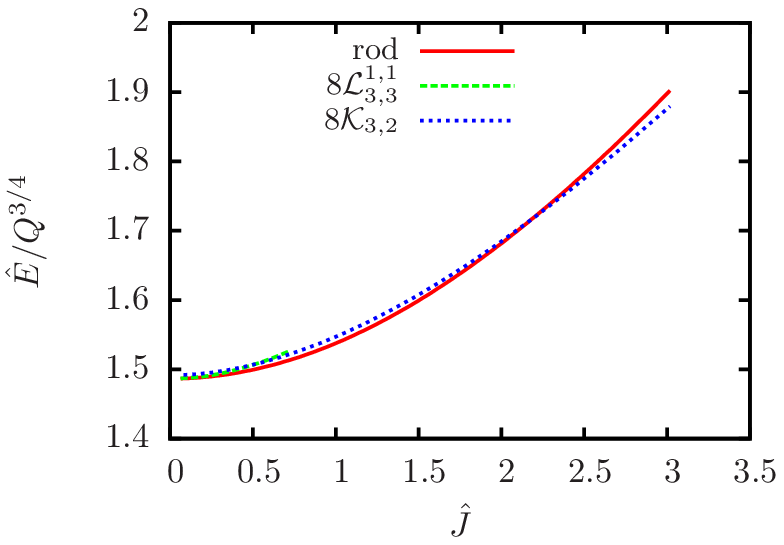}
}
   \caption{Solitons with $Q=8$ and $\mu=2$.  The top row shows position curves at $\omega =0$ (light green) and $\omega=1$ (blue).  The bottom row shows energy as a function of the parameter $\omega$ and as a function of isospin $J$.}
   \label{fig10}
\end{figure}

At degree five we found two distinct local minima of the pseudo-energy, namely a $5\mathcal{L}_{1,1}^{1,2}$ link and a $5\widetilde{\mathcal{A}}_{5,1}$ buckled ring, see figure \ref{fig7}.  When $\omega=0$ the link has the lower energy.  The two curves for $E(\omega)$ cross when $\omega=0.82$ and thereafter the ring has the lower energy.  In contrast, the $E(J)$ curves for these configurations do not cross and the link has the lower energy for any given value of $J$.

At degree six we again found two distinct local minima of the pseudo-energy.  These were links of type $6\mathcal{L}_{2,2}^{1,1}$ and $6\mathcal{L}_{3,1}^{1,1}$, shown in figure \ref{fig8}.  Unlike the study \cite{fos}, we did not find a $6\mathcal{A}_{3,2}$ configuration; this could be due to a different choice of potential function or a different choice of the mass parameter $\mu$.  The $6\mathcal{L}_{2,2}^{1,1}$ link has a lower energy than the $6\mathcal{L}_{3,1}^{1,1}$ link for all values of $\omega$ and for all values of $J$.

At degree seven the only minimum of the pseudo-energy found was a $7\mathcal{K}_{3,2}$ knot, and this is shown in figure \ref{fig9}.

At degree eight the three energy minima found were a link of type $8\mathcal{L}_{3,3}^{1,1}$ and a knot of type $8\mathcal{K}_{3,2}$, see figure \ref{fig10}.  We did not find a soliton corresponding to the $8\mathcal{A}_{4,2}$ configuration in \cite{fos}.
Within the limits of numerical accuracy the $8\mathcal{L}_{3,3}^{1,1}$ and $8\mathcal{K}_{3,2}$ configurations are degenerate in energy when $\omega=0$.  As $\omega$ increases the knot energy grows faster than that of the link, so that the link has the lower energy.  When $\omega$ reaches the value 0.38 the link collapses to the knot, which has a smaller pseudo-energy.  Our algorithm is unable to find the link at larger values of $\omega$.  It is likely that when $\omega\geq0.38$ the link is an unstable critical point of the pseudo-energy, and that it continues to have a lower energy than the knot.  As far as we can tell, the $E(J)$ curves for the two solitons coincide, at least within the range of values of $J$ in which both can be constructed. Note that, since their energies are
indistinguishable at $\omega=0$, this is precisely what the rod model
predicts.

\section{Conclusions}
\label{sec:con}
\news

To summarize, we have reformulated the
 problem of constructing internally spinning
Hopf solitons of frequency $\omega$
as a variational problem for pseudoenergy $F_\omega$,
and solved this problem numerically for Hopf charges 1--8, 
$0\leq \omega\leq 1$ without
imposing any spatial symmetries. For each local minimizer of the
static energy $F_0$ we have constructed a curve of isospinning hopfions,
parametrized by $\omega$, for which the total energy
$E(\omega)$ and conserved isospin $J(\omega)$ were calculated. 
Generically, the solitons persist for all $\omega\leq\min\{1,\mu\}$, where
$\mu$ is the meson mass of the model, and their qualitative shape is
independent of $\omega$. We noted two exceptions to this: for $Q=2$, $\mu=1$,
the solitons undergo a bifurcation as $\omega$ increases through
0.86, branching from $2{\cal A}_{2,1}$ to $2{\cal \wt{A}}_{2,1}$ (i.e.\ they
lose axial symmetry); for $Q=4$, $\mu=2$,
 the axially symmetric $4{\cal A}_{2,2}$
soliton branches to a link $4{\cal L}_{1,1}^{1,1}$ as $\omega$ increases
through 0.58. Even when no such bifurcation occurs, the shape of the lowest
energy soliton with fixed $\omega$ can change as $\omega$ varies, 
because the graphs of $E(\omega)$ for different solitons with the same 
Hopf degree can cross. We found that this happens for $Q=5$, $\mu=2$, for
example. Similarly, the shape of the lowest energy soliton with fixed
isospin $J$ can change as $J$ varies, if the graphs of $E(J)$
for different soliton branches cross, as happens, for example for $Q=4$, 
$\mu=2$. Such crossings seem to be comparatively rare, however.

We have developed a time-dependent extension of the the elastic rod model
of Hopf solitons, which predicts simple, uniform scaling behaviour of
the total energy and isospin of hopfions as $\omega$ varies, namely
\beq
E(\omega)=\frac{E_0}{\sqrt{1-(\omega/\omega_*)^2}},\qquad
J(\omega)=\frac{(\omega/\omega_*^2)E_0}{\sqrt{1-(\omega/\omega_*)^2}}
\eeq
where $E_0$ is the static soliton energy and $\omega_*$ is an (unknown)
parameter, depending only on the details of the Faddeev-Skyrme model
under consideration (but independent of $Q$, and of which branch of local
minimizers one is following). Despite having only one free parameter, the
rod model fits the $E(\omega)$ graphs well for small $\omega$,
and the $E(J)$ graphs well for all $J$. The rod model predicts
that the energy graphs never cross, and that solitons never change shape
(they just dilate uniformly), as $\omega$ (or $J$) varies which, again, is
in good qualitative agreement with our numerical results.

\section*{Acknowledgments}

This work was financially supported by the UK Engineering and 
Physical 
Sciences Research Council (grant number EP/G009678/1). Some of the work of
JJ was undertaken at the University of Antwerp, Belgium, financially
supported by an FWO Visiting Post-Doctoral Fellowship, and some of the 
work of DH was carried out at Durham University, supported by the Engineering 
and Physical Sciences Research Council (grant number EP/G038775/1). 
Most numerical calculations were performed on the
ARC1 HPC  system  at the University of Leeds, some 
 on the cluster SKIF at the Belarusian State University. 
We would like to thank  Paul Sutcliffe, Steffen Krusch, David Foster, 
Alexey Halawanau and Joan Camps for helpful discussions. 

\subsection*{Note added in proof}

Similar numerical results to those presented here were obtained independently
by Haberichter and Battye, in a preprint which appeared shortly after 
ours \cite{bathab}. They use a similar numerical strategy, but with
damped field evolution instead of the quasi-Newton BFGS method to implement
gradient descent. 
 There are minor differences in some results, which are probably due to
a different choice of potential, namely $U(\varphi)=\mu^2(1-\varphi_3)$, 
instead of our choice $U(\varphi)=\frac12\mu^2(1-\varphi_3^2)$.''

\end{document}